\documentclass[graybox]{svmult}

\usepackage{type1cm}        
\usepackage{makeidx}         
\usepackage{graphicx}        
\usepackage{multicol}        
\usepackage[bottom]{footmisc}

\usepackage{newtxtext}       %
\usepackage{newtxmath}       

\usepackage[colorlinks=false, linktocpage=true,backref,pagebackref,hidelinks]{hyperref}
\usepackage{cprotect}

\makeindex             

\newcommand{\R}{\mathbb R}

\newcommand{\M}{\mathbb M}

\newcommand{\beq}{\begin{equation}}
\newcommand{\eeq}{\end{equation}}
\newcommand{\beqarr}{\begin{eqnarray}}
\newcommand{\eeqarr}{\end{eqnarray}}
\newcommand{\beqa}{\begin{eqnarray*}}
\newcommand{\eeqa}{\end{eqnarray*}}
\usepackage{natbib}

\begin{document}
\setcounter{chapter}{98}

\title*{Gauging the spacetime metric --
{ looking back and
forth a century later}  {\tt \normalsize $\quad$ version 01/11/2019} }
\author{Erhard Scholz}
\institute{University of Wuppertal, Faculty of  Math./Natural Sciences, and Interdisciplinary Centre for History and Philosophy of Science, \email{scholz@math.uni-wuppertal.de}
}
%
%

\maketitle

\abstract{H. Weyl's proposal of 1918 for generalizing Riemannian geometry by  local scale gauge (later called {\em Weyl geometry}) was motivated  by mathematical, philosophical and physical considerations. It was the starting point of his unified field theory of electromagnetism and gravity. After getting disillusioned with this research program  and after the rise of  a convincing alternative for the gauge idea  by translating it to the phase of  wave functions and spinor fields in quantum mechanics, Weyl no longer considered the original scale gauge  as  physically relevant.
   About the middle of the last century the question of conformal and/or local scale gauge transformation were reconsidered by different authors  in high energy physics (Bopp, Wess, et al.)  and, independently, in gravitation theory (Jordan, Fierz, Brans, Dicke).  In this    context Weyl geometry attracted new interest among different groups of physicists  (Omote/Utiyama/Kugo, Dirac/Canuto/Maeder, Ehlers/Pirani/Schild and others), often by hypothesizing a new scalar field linked to gravity and/or high energy physics.  Although not crowned by immediate success, this ``retake'' of Weyl geometrical methods  lives on and has been extended a century after Weyl's first proposal of his basic geometrical  structure. It finds new interest in present day studies of elementary particle physics, cosmology, and philosophy of physics. 
  }

\tableofcontents

\newpage
\section{Introduction}
\label{Intro}
In 1918 {\em  Hermann Weyl} proposed a generalization of Riemannian geometry, which he considered as better  adapted to the field theoretic context of general relativity than the latter. His declared intention was to base geometry   on purely ``local'' concepts which at the outset would not allow to compare  field quantities  $X(p)$ and $X(p')$ at distant points $p$ and $p'$  of the spacetime manifold, but only for infinitesimally close ones.  The possibility of comparing  directly two quantities at  distant points appeared to him a remnant of  Euclidean  geometry, which Riemannian geometry had inherited via the Gaussian theory of surfaces. In Weyl's view Riemann  had  generalized the latter without putting the comparability of quantities at different locations into question. He  demanded in contrast that

\begin{quote}
\ldots  only segments at the same place can be measured against each other. The gauging of segments is to be carried out at  each single place of the world (Weltstelle), this task {\em cannot be delegated to a central office of standards (zentrales Eichamt)}. \citep[p. 56f., emph. ES]{Weyl:Erweiterung}\footnote{``\ldots nur Strecken, die sich an der gleichen Stelle befinden, lassen sich aneinander messen. An jeder einzelnen Weltstelle mu{\ss} die Streckeneichung vorgenommen werden, diese Aufgabe kann nicht einem zentralen Eichamt \"ubertragen werden'' \citep[p. 56f.]{Weyl:Erweiterung}. (Translation here and in the following by ES, if no reference to a published English translation is given.)}
\end{quote}

He therefore considered a  geometry  formalized by a conformal (pseudo-Riemannian) structure as more fundamental than  Riemannian geometry itself  \citep[p. 13]{Weyl:InfGeo}. But it has to be supplemented by a principle which allows  for comparing   metrical quantities  at infinitesimally close points ($p \neq p'$),  realized by  a principle of metrical transfer. A conformal manifold could be qualified as ``metrically connected'' only if a comparison of metrical  quantities at different points is possible:
\begin{quote}
A metrical connection from point to point is only being introduced into it [the manifold, ES], if a {\em principle of transfer for the unit of length from one point $P$ to an infinitesimally close one} is given. \citep[14]{Weyl:InfGeo}\footnote{Emphasis  in the original, here and in other places, where  not explicitly stated that it is by ES.} 
\end{quote}
Weyl formulated this principle of transfer by introducing what today would be called a connection in the local scaling group $\R^+$, i.e., by  a real differential 1-form. The new type of ``purely infinitesimal geometry'' (Weyl's terminology), later called {\em Weyl geometry}, was built upon the two interrelated basic concepts of a {\em conformal structure} and a {\em  scale connection} as the principle of transfer. Weyl called the latter a {\em length connection}.  Both were united by the possibility  of changing the metrical scale by {\em gauge transformations} in the literal sense (see sec. \ref{subsec: Inf Geo}). For a few years Weyl tried to build a  unified field theory of electromagnetism and gravity upon such a geometrical structure,  and to extend it to a field theory of matter.\footnote{Cf. \citep{Vizgin:UFT,Goenner:UFT}.}
 But in the second half of the 1920s he accepted and even contributed actively to reformulating the gauge idea in the context of the rising new quantum mechanics.
 
Several decades later this idea was generalized to non-abelian groups and became a fundamental conceptual ingredient even for the lager development of high energy physics.\footnote{See N. Straumann's contribution to this volume.}
 In the meantime  (during the 1940s) Weyl had  recanted  the importance of scaling transformations (localized ``similarities'' as he used to call them) for the search of what he called the ``physical automorphisms of the world'' (see sec. \ref{subsec: Withdrawel}).

The basic idea underlying Weyl's ``purely infinitesimal geometry'' of 1918 reappeared independently at several occasions during  the second half of the 20th century. In the early 1960s  Carl Brans and Robert Dicke developed their program of a modified general relativistic theory of gravity with a non-minimally coupled scalar field.  Dicke stated as an ``evident'' principle (which it was  {\em not}, at least not for everyone):
\begin{quote}
It is evident that the particular values of the units of mass, length, and time employed are arbitrary and that the
{\em laws of physics must be invariant under a general coordinate dependent change of units}. \citep[p. 2163, emph. ES]{Dicke:1962} \label{Dicke-evident}
\end{quote}
This was very close to Weyl's view in 1918, but Dicke postulated local scale invariance without the complementary structure of a scale connection. As a result Brans, Dicke's PhD student, and Dicke himself developed a theory of gravity which had an  implicit relationship to the special type of Weyl geometry with an integrable scale connection, in short {\em integrable Weyl geometry} (IWG). Other authors made this relationship explicit and generalized it to the non-integrable case.

This was part of a classical field theory  program of gravity research, but the importance of conformal transformations got also new input from high energy physics. And even the original form of Weyl geometry had some kind of revival from the 1960s onward in scalar field theories of gravity or nuclear structures, initiated by authors in Japan (Omote/Utiyama/Kugo) and, independently Europe/USA  (Dirac/Canuto/Maeder),  and also in the foundational studies of gravity (Ehlers/Pirani/Schild). This restart in the last third of the 20th century of  research building on, and extending, Weyl geometric methods in physics has lasted until today and shows that Weyl's disassociation from his scale gauge idea is not at all to be considered  a final verdict on his geometrical ideas developed between  1918 and  the early 1920s.  

The following paper tries to give an account of the century long development  from Weyl's  original scale gauge geometry of 1918 (and the reasons why he thought it an important improvement on the earlier field theories building upon Riemannian geometry),  through its temporary disregard induced by the migration of the gauge idea from metrical scale to quantum phase  (ca. 1930--1960, section  \ref{sec: Weyl geometry 1918}),  and the  revival in the early 1970s indicated above (section \ref{sec: Retake}), 
to a report on selected research in physics, which uses Weyl geometric methods in a crucial way (section  \ref{sec: Today}). 
Basic concepts and notations  of Weyl geometric gravity (in the moderately modernized form  in which they are used in  section   \ref{sec: Today})  are  explained  in an interlude  between the   historical account  and the survey of present studies (sec. \ref{sec: Interlude}).  Short reflections on this glance back and forth are given at the end of the paper (section \ref{sec: Looking back and forth}).


\section{Weyl's scale gauge geometry  1918--1930}
\label{sec: Weyl geometry 1918}

\subsection{Purely infinitesimal geometry, 1918-1923 }
\label{subsec: Inf Geo}
Parallel to finalizing his book {\em Raum - Zeit - Materie} (RZM) \citep{Weyl:RZM} Weyl developed a generalization of Riemannian geometry, in which  an inbuilt concept for a direct metrical comparison of quantities at distant points was no longer foreseen. It was substituted by a comparison in ``purely local'' regions, in the infinitesimal sense   \citep{Weyl:GuE,Weyl:InfGeo}. Weyl introduced this generalization into the third and fourth editions of the book,  and  added a discussion of what might be the consequences for  relativistic field theory and the description of matter. In this way his proposal entered the English translation {\em Space - Time - Matter} (STM) of the fourth edition   of the book \citep{Weyl:STM}. It became more widely known than his separate articles on the topic.  

In a  letter dated 1 March 1918 to {\em Albert Einstein}, before RZM was publicly available,  Weyl announced that the publisher (Springer) would soon send the corrected page proofs of the book to Einstein.\footnote{\citep[vol. 8, doc 472]{Einstein:CP}}
 In March the two men met in Berlin, and Weyl used the occasion for presenting his ideas on the generalized theory  to Einstein.  This started a friendly, but controversial discussion on Weyl's proposals which would extend through the whole year 1918  and continued to have some repercussions  in the  years to come.\footnote{\citep{Straumann:Einstein_Weyl,Lehmkuhl:Einstein-Weyl}, all the letters in  \citep[vol. 8B]{Einstein:CP}.}    
 Weyl submitted his first publication on the topic to the Berlin Academy of Science through  Einstein who appended a famous critical note to it  \citep{Weyl:GuE}. As Einstein queried, atomic clocks (spectral lines  of atoms/molecules) would become dependent on their  history, if one assumes that their internal time  is subject to Weyl's local length transfer.  Weyl did not share this assumption; he considered the length connection as a part of the general field theoretic structure, which is reflected only indirectly in the behaviour of measuring instruments. 
 This  was one difference among others which the two scientists debated in the next few  years.\footnote{The  debate between Einstein and Weyl after 1918 is being dealt with in  \citep{Lehmkuhl:Einstein-Weyl}.}
 
Other  physicists, among them {\em Arthur Eddington} and the young {\em Wolfgang Pauli}, reacted differently. For a period of a few years they contributed  to the dissemination and an elaboration of Weyl's theory. The latter  might have been of particular interest to physicists at the time,  because it seemed to  open a geometrical road towards a unification of the then known fundamental interactions, electromagnetism and gravity, and was related to the even more ambitious goal of a fundamental theory of matter based on a unified classical field theory along the lines of G. Mie and D. Hilbert. 

A few years later this early program of a geometrical unification became considered obsolete, when the gauge idea was transformed from  geometrical scale to the quantum mechanical phase degree of freedom. In this form it was the starting point for the  gauge field theories of the second half of the last century, which became central  for the standard model of particle physics.  The early parts of this  interesting  story has been  told from  different angles;\footnote{See, among others, \citep{Vizgin:UFT,Sigurdsson:Diss,Goenner:UFT,Scholz:Weyl_matter,%
Ryckman:Relativity,Afriat:2013,Afriat:2019}.} 
it will not be repeated here. The later part -- the rise of the standard model -- would probably still deserve more detailed historical investigations.\footnote{See  \citep[chap. 22]{Kragh:Quantum}  \citep{Pickering:Quarks,Galison:Image-Logic,Borrelli:isospin,Borrelli:2015,Borrelli:Higgs}}  Here we concentrate  on Weyl's basic conceptual ideas for gauging the metric of spacetime, which lay at the basis of his new geometry. Not unlike the more general idea of gauge structures (with groups operating on dynamical spaces of field variables)  it  proved of a wider range and impact than Weyl's first approach for a physical application in unified field and matter theory. 

 Weyl considered the {\em orthogonality} relation specified by a symmetric, non-degenerate (but not necessarily definite) bilinear form in the coordinate differentials  as an  {\em objective} element for introducing a  {\em metric} (``Ma{\ss}bestimmung'') in a differentiable manifold $M$. It is given  independently of any further choice of the description; for a Lorentzian signature it corresponds to the causal structure of spacetime. On the other hand,  the  scale factor (``Proportionalit\"atsfaktor'') of the bilinear form is  part of  the choice of a {\em reference frame} (``Bezugssystem'') and  defines a metrical {\em  gauge}   (``Eichung'')  in the literal sense \citep[p. 58]{Weyl:Erweiterung}.\footnote{
 Page references here and in the following refer to Weyl's {\em Gesammelte Abhandlungen}.} In this sense it is a  subjective element of the description like the choice of a coordinate system   and complementary to it \citep[p. 13]{Weyl:InfGeo}.
 
  Thus far $M$ carries a {\em conformal}  structure, or in Weyl's words a ``conformal geometry'' (ibid.).\footnote{The terminology of different {\em structures} on a manifold appeared only in the 1940s, essentially due to protagonists of the French community (C. Ehresmann et al.).} 
But this is insufficient for establishing a full-fledged metrical concept in the manifold. Without further stipulations the different points of the manifold would ``be completely isolated from each other from the metrical point of view''. Weyl concluded:
\begin{quote}
A metrical connection between points is being introduced only 
if  a {\em principle of transfer for the unit of length measurement from a point $P$   to an infinitesimally close one } is given. \citep[p.14]{Weyl:InfGeo}\footnote{``Machen wir keine weiteren Voraussetzugen, so bleiben die einzelnen Punkte der Mannigfaltigkeit in metrischer Hinsicht vollst\"andig gegeneinander isoliert. Ein metrischer Zusammenhang wird erst dann in sie hineingetragen, wenn ein {\em Prinzip der \"Ubertragung der L\"angeneinheit von einem Punkt $P$ zu einem unendlich benachbarten} vorliegt. \citep[p. 14]{Weyl:InfGeo} }
\end{quote}

He  specified such a principle of transfer by adding  a differential 1-form $\varphi = \varphi_k dx^k$ to the metrical coefficients $g_{ik}$ of the chosen reference frame.\footnote{In fact, Weyl wrote ``$d\varphi = \varphi_i dx_i$'' for the differential form, e.g. \citep[p. 15]{Weyl:InfGeo}. In order to avoid confusion with the present notational conventions  influenced by Cartan's symbolism of differential forms,  we denote it by  $\varphi$ because  in general it is not an exact form.}
For him the differential form codified the change of squared length quantities $l^2= l^2(p)$  measured at a point $p\in M$ to the respective squared lengths $l'^2 = l^2(p')$ measured at a infinitesimally close point $p'$ expressed by  the relation
\beq l'^2 = (1+\varphi(\xi)) \, l^2 \; .
\eeq 
Here the infinitesimal displacement of the points isdenoted in terms of a tangent vector $\xi \in T_pM$. 
But then, of course, one has to gain clarity about the  transformation $\varphi$ undergoes if the local scale is changed, i.e. if the metric  $g_{\mu \nu}$  is conformally  changed to $\tilde{g}_{\mu\nu}=\lambda(x) g_{\mu \nu}$. Weyl realized that  consistency of the length transfer idea demands to transform the scale connection $\varphi$   as follows:
\beq \mbox{If}\quad g\; \longmapsto \tilde{g}=\lambda g\, ,  \quad \mbox{then} \quad \varphi \; \longmapsto \; \tilde{\varphi} = \varphi - \frac{1}{2} d \log \lambda \, , \;\;  \label{eq gauge trafo}
\eeq
i.e., $\tilde{\varphi}_{\mu} = \varphi_{\mu} - \frac{1}{2}\lambda^{-1}\partial_{\mu}\lambda $. 
 Weyl called it a ``modification of the  gauge (Ab\"anderung der Eichung)''  \citep[p. 59]{Weyl:Erweiterung};  this was the first {\em gauge transformation} considered explicitly  in the history of mathematics/physics.
  
He showed that  a genuine geometry on a manifold $M$ can be built upon such a generalized concept of a gauge metric (later {\em Weylian metric}). Slightly reformulated, Weyl characterized his  metric  by an  equivalence class $[(g, \varphi)]$ of pairs $(g,\varphi)$ consisting of of a (pseudo-) Riemannian metric $g$ on $M$ and a real valued differential form $\varphi$ on $M$, equivalences being  given by  gauge transformations (\ref{eq gauge trafo}).
Gauging the metric boils down to choosing a representative  $(g,\varphi)$; its  first component  $g$ will be called the {\em Riemannian component} of the respective gauge, the second one $\varphi$ its {\em scale connection form}.
 
A clue to this geometry was Weyl's finding that  $[(g,\varphi)]$  possesses a {\em uniquely determined} (in particular gauge independent), {\em metric compatible, affine connection} $\Gamma$ (in the sense of a symmetric linear connection)  like in Riemannian geometry \citep[p. 33]{Weyl:GuE}.\footnote{In \citep[p. 17]{Weyl:InfGeo} Weyl referred also to \citep{Hessenberg:1917} in this regard.}  Metric ``compatibility'' is here understood as the consistency of the length change of vectors under parallel transport by $\Gamma$ with the length change demanded by the length transfer $\varphi$. $\Gamma$ is gauge independent, although in any gauge $(g,\varphi)$ it can be decomposed into two gauge dependent terms, the Levi-Civita connection $\Gamma(g)$ of $g$ and a scale connection term $\Gamma = \Gamma(\varphi)$, 
\beq \Gamma = \Gamma (g)+ \Gamma (\varphi)\,  \quad \mbox{where} \quad  \Gamma^{\mu}_{\nu \lambda}(\varphi) = 
\frac{1}{2}( \delta^{\mu}_{\nu} \varphi_{\lambda} +  \delta^{\mu}_{\lambda} \varphi_{\nu} - g_{\nu \lambda}\,\varphi^{\mu}) \, . \label{eq affine connection}
\eeq  
$\Gamma$  induces a gauge independent {\em covariant derivative} and has geodesics which are no longer extremal lines of the (gauged) metric but are  autoparallels of the invariant affine connection.

Weyl derived the {\em curvature} tensors for the connections $\Gamma$ and  $\varphi$. With regard to the affine connection he formed  the Weyl geometric curvature tensor  analogous to the Riemann tensor in the classical case. Because of this analogy it will be  denoted   in the following  by $\mathit{Riem}$  (with components $R^{\mu}_{\;\; \nu \lambda \kappa}$) although it (more precisely its $(0,4)$ type version  $R_{\mu \nu \lambda \kappa}$) is not antisymmetric in the first pair of indices. Its contraction $\mathit{Ric}$ is  the Weyl geometric Ricci tensor, and   $R$ its scalar curvature. Like $\Gamma$, the Weyl geometric curvatures can be decomposed, in any gauge $(g,\varphi)$, into contributions derived from the Riemannian part of the metric $g_{\mu\nu}$ alone, the Riemannian part of the curvature $\mathit{Riem}\, (g) = \mathit{Riem}\,(\Gamma(g))$,  and the scale connection part of the curvature $\mathit{Riem}\, (\varphi)$, which 
 is formally built like the usual Riemann tensor expression  from $\Gamma (\varphi)$,  but with covariant derivatives $_g\hspace{-0.2em}\nabla_{\kappa}$ of $g$ in place of the partial derivatives  $\partial_{\kappa}$.\footnote{Cf. footnote \ref{fn curvature conventions}; more details in \citep{Yuan/Huang:2013} or in \citep[p. 218f.]{Eddington:Relativity},   \citep[p. 150ff.]{Perlick:Diss} etc.}

The curvature formed with regard to the scale (length) connection is $f= d \varphi$, written in Cartan notation of outer derivatives (which Weyl did not use). It is called the {\em length (scale) curvature}.  Weyl decomposed the curvature tensor $Riem$  into a contribution $\overline{R}^{\,\mu}_{\,\;\; \nu \lambda \kappa}$  with the same symmetry properties as the classical Riemann tensor  and a contribution derived from the length curvature, which is not antisymmetric in the first two indices:
\beq R^{\mu}_{\;\; \nu \lambda \kappa} = \, \overline{R}^{\,\mu}_{\,\;\; \nu \lambda \kappa} - \frac{1}{2}\delta^{\mu}_{\nu}f_{ \lambda \kappa} \;
\eeq
Also this decomposition is gauge independent.  Weyl called the first term the {\em directional curvature} (``Richtungskr\"ummung'') \citep[p. 20]{Weyl:InfGeo}. 

From the directional curvature he constructed a  tensor $C$ of type $(1,3)$ like $Riem$, which ``depends only on the $g_{jk}$''. It is   conformally invariant and vanishes in the dimensions $n=2,3$. For $n\geq 4$, so  Weyl announced, it vanishes if the manifold is  conformally flat \citep[p. 21]{Weyl:InfGeo}. Later it would be called  {\em conformal curvature} or  the {\em Weyl tensor}.\footnote{That  $C=0$ is also  {\em sufficient} for conformal  flatness was not clear to Weyl in 1918; he even seemed to exclude it in a side remark. It was proven by \citet{Schouten:1921} and later by himself  \citep{Weyl:projektiv_konform}. }

For vanishing length curvature, $f=d \varphi=0$, the scale connection can be  integrated locally. Then the corresponding gauge reduces to  a Riemannian metric; it will therefore be called the {\em Riemann gauge} of the   Weylian metric. In this case the curvature tensor reduces to the directional curvature and is identical to the Riemann curvature of the Riemann gauge. Weyl therefore considered Riemannian geometry as a special case of his scale gauge geometry. This was his perspective in RZM from the third edition onward and also in its English version STM \citep[chap. 2]{Weyl:STM}.

The generalization of the metrical structure demanded an extension of tensor (of course also of vector and scalar) fields with regard to their scaling properties under a change of metrical gauge. In the third edition of RZM Weyl wrote
\begin{quote}
In a generalised sense we shall, however, also
call a linear form which depends on the co-ordinate system and the
{\em calibration} a tensor, if it is transformed in the usual way when
the co-ordinate system is changed, but becomes multiplied by the
factor $\lambda^e$ (where $\lambda=$ the calibration ratio) when the calibration is
changed; we say that it is of {\em weight} $e$. \citep[p. 127]{Weyl:STM}\footnote{In the 5th  edition \citep[p. 127]{Weyl:RZM5}  }
\end{quote}
 Weyl considered these fields ``merely as
an auxiliary conception, which is introduced to simplify calculations'' (ibid.); but  physicists among his readers  realized that it would become of physical importance if Weyl geometry is accepted as a framework for field theory. Einstein called them ``Weyl tensors'' \citep[p. 200]{Einstein:Ergaenzung},  Eddington introduced the terminology of ``co-tensors'' or ``co-invariants''  \citep[p. 202]{Eddington:Relativity}, which is still  in use in parts of the physical literature. Here they will be called {\em scale covariant} quantities (tensor, vector, scalar fields).

Weyl's conjectural unification of gravity and electromagnetism built essentially on the idea of using the length connection $\varphi$ as a symbolic representative of the electromagnetic potential. Then the Maxwell action density  $ f_{\nu \lambda}f^{\nu \lambda} \sqrt{|g|}$ for the electromagnetic field was of weight 0,  if and only if the dimension is $n=4$. This was an intriguing argument for the necessary specification of the spacetime dimension 4 in Weyl geometric field theory \citep[p. 31]{Weyl:GuE}, \citep[p. 37]{Weyl:GuE_English}. In an early response to Weyl's theory it was praised by Einstein  also.\footnote{``Ihr Gedankengang ist von wunderbarer Geschlossenheit. Auch der Schluss auf die Dimensions\-zahl 4 hat mir sehr imponiert.'' \citep[vol. 8B, Doc 499, 8 Apr. 1918]{Einstein:CP}.  Weyl could not keep up this argument after 1927; for his later deliberations see \citep{DeBianchi:2019}.}
 In order to get a  scale invariant gravitational action density, Weyl replaced the Hilbert term $ R \sqrt|g|$ by  quadratic expressions in the Weyl geometric curvature, essentially $R^2 \sqrt|g|$, also of weight $0$. 

In such an approach the Weylian metric seemed to establish a unified description of gravity ($g_{\mu \nu}$ and affine connection $\Gamma$)  and electromagnetism ($\varphi$ and its curvature $f=d\varphi$). But the question of the physical interpretation of the scale invariant geodesics and its relation to the free fall trajectories of neutral and of charged particles had to be answered. This was part of the discussion between Weyl and Einstein, aside from Einstein's spectral line objection to Weyl's geometrical generalization.  For Weyl the latter was no compelling (knock-out) argument. He answered that one has to distinguish between the field theoretic metric and the measure indicated by clocks and rods. The latter would be realized by a specific gauge and ought to depend on the physical behaviour of atomic systems in the local (infinitesimal) neighbourhood of the atom, not on their history. 

 For the time being he proposed to consider the hypothesis that atoms adapt to the local field constellation of the gravitational field in such a manner that the Weyl geometric scalar curvature becomes constant,  $R = \mathit{const}$ \citep[p. 298f.]{Weyl:RZM5}.\footnote{Weyl  mentioned this argument already in his letter  of 18 Sep. 1918 to Einstein \citep[vol. 8B, Doc. 619, p. 877]{Einstein:CP}, but  introduced it into RZM only in the 5th edition; it is thus not contained  in  the English translation \citep{Weyl:STM}. The Weyl geometric scalar curvature $R$  scales with the inverse metric $g^{\mu \nu}$. If it is different from zero it can thus be scaled to a constant like any other non vanishing scalar field of the same scale weight.}
 \label{Weyl clock adaptation}
At the time of the fifth edition of RZM (1923) he still kept to the position that the final judgement on his theory would  depend on a theory of measurement.  Three years later he gave up this idea and joined the interpretation of the electromagnetic  potential as a phase connection in the new quantum mechanics.

In the two  years before the fifth edition, on the other hand, Weyl looked for a deeper philosophical-conceptual underpinning of his geometry in  a new {\em analysis of the problem of space} (PoS). Half a century earlier Helmholtz had analysed the principles which the motions of a rigid body have to satisfy in order for being able to establish empirical measuring rules. The evaluation of these principles led to  classical Euclidean or non-Euclidean geometry.  Soon Helmholtz's principles for free mobility of rigid rods were rephrased in terms of conditions for the automorphism group of space.\footnote{Cf. \citep{Merker:Espace,Bernard:2018Paris}. At first Helmholtz believed to have derived Euclidean geometry alone, but soon learned that non-Euclidean geometry  satisfies his principles too. }
 This analysis became known as the classical PoS.

The fusion of space and time in a  unified spacetime in special relativity and the loss of the concept of rigid body in the general theory  let Helmholtz's analysis appear obsolete in the light of relativity. At the end of 1920   Weyl set out for formulating conceptual principles for any type of  geometry in a smooth manifold,  which builds  upon congruence and similarity operations in the infinitesimal neighbourhoods only. As a unifying (``synthetical'')  principle he included the postulate of a {\em  uniquely determined affine connection} and formulated the result  of his conceptual analysis  in the form of axioms for those  subgroups of the general linear group, which may serve as candidates for defining the local similarity and congruence relations in the manifold in the mentioned sense  \citep{Weyl:ARP1921,Weyl:ARP1922b}. In 1922 and early  1923 he was able to show that the only groups  satisfying his axioms in a manifold of dimension $n$ are the special pseudo-orthogonal groups $SO(p,q;\R)$  as ``congruence'' groups ($ p+q=n$), with  similarities  $SO(p,g;\R) \times \R^+$
 \citep{ Weyl:ARP1923}.  If one accepts his characterization of the  principles for infinitesimal congruence and similarity relations this can be read as a strong conceptual underpinning for the structure of Weylian manifolds.\footnote{More details in \citep{Bernard:Barcelona,Bernard/Lobo:PoS,Scholz:2016Weyl/Cartan}. A short version of the argument can also be found in the fifth edition of RZM \citep{Weyl:RZM5}, a provisional sketch already in the 4th edition and in \citep{Weyl:STM}.}

Another conceptual insight of long range was gained by Weyl in a paper  which arose  as a by-product to a report he wrote for F. Klein on a manuscript by J.A. Schouten  \citep{Weyl:projektiv_konform}. The manuscript had been rejected by L.E.J Brouwer for the {\em Mathematische Annalen}, but was published a little later in {\em Mathematische Zeitschrift} \citep{Schouten:1921}.\footnote{Klein to Weyl 6.10.1920, Weyl to Klein 28.12.1920, University Library G\"ottingen, Codex Ms Klein 296, 297.} 
Schouten showed that  in dimension $n>3$ the vanishing of the conformal tensor $C$ of a Riemannian metric $g_{\mu \nu}$ implies local conformal  flatness. For Weyl this was new and gave him an incentive to study the projective and conformal viewpoint which he had  ``touched upon only marginally'' in his previous discussions of infinitesimal geometry \citep[p. 201]{Weyl:projektiv_konform}. He started the paper with the remark:
\begin{quote}
The construction of pure infinitesimal geometry, which I have described most consequentially in the 3rd and 4th edition of my book {\em Raum, Zeit, Materie}, is naturally implemented in the three levels which are characterized by the catchwords {\em continuous connection, affine connection, metric}. {\em Projective} and {\em conformal} geometry originate by means of abstraction from the affine respectively the metric one. \citep[p. 195]{Weyl:projektiv_konform}\footnote{``Der Aufbau der reinen Infinitesimalgeometrie, wie ich ihm am folgerichtigsten in der 3. und 4. Auflage meine Buches {\em Raum, Zeit, Materie} geschildert habe, vollzieht sich nat\"urlicherweise in drei Stockwerken, welche durch die Schlagworte {\em stetiger Zusammenhang, affiner Zusammenhang, Metrik} gekennzeichnet sind.  Die {\em projektive} und {\em konforme} Geometrie enspringen durch Abstraktion aus der affinen bzw. der metrischen.'' \citep[p. 195]{Weyl:projektiv_konform}  
}
\end{quote}
 Under  ``continuous connection (stetiger Zusammenhang)'' Weyl understood the possibility to consider infinitesimal neighbourhoods; in present terms it is usually expressed  as a smooth structure on the underlying manifold.\footnote{For a philosophical analysis of this idea see \citep{Bernard:Idealisme}. A modern reconstruction which may even be closer to Weyl's intentions than their reformulation in terms of standard differential topology,  may be possible by differential geometry with infinitesimals; see \citep{Kock:2006}. }
 
  Weyl defined the ``projective property (projektive Beschaffenheit)'', in our terms the {\em projective structure} of a smooth manifold $M$, by means of an equivalence class $[\Gamma]_p$ of affine connections $\Gamma$ which have ``geodesics'' (autoparallels) with identical traces. $\Gamma$ and $\overline{\Gamma}$ are projectively equivalent, if 
 \beq \overline{\Gamma}^{\mu}_{\nu \lambda} = {\Gamma}^{\mu}_{\nu \lambda} + \delta^{\mu}_{\nu} \psi_{\lambda} + \delta^{\mu}_{\lambda} \psi_{\nu}  \label{eq Gamma}
 \eeq 
for some differential form $\psi =\psi_{\nu}dx^{\nu}$. 
He characterized the {\em conformal structure} (``konforme Beschaffenheit'') of $M$ analogously by an equivalence class of affine connections  $[\Gamma]_{c}$.  Here he considered two 
affine connections $\Gamma$ and $\widetilde{\Gamma}$ as belonging to the same  conformal class  $[\Gamma]_{c}$, if both are scale invariant affine connections of  Weylian metrics $[(g,\varphi)]$ and $[(\tilde{g}, \tilde{\varphi})]$ with conformal Riemannian components, $g\sim \tilde{g}$. Because of (\ref{eq affine connection}) this boils down to the condition \citep[p. 196]{Weyl:projektiv_konform}
\beq \widetilde{\Gamma}^{\mu}_{\nu \lambda} - {\Gamma}^{\mu}_{\nu \lambda}  = \frac{1}{2}( \delta^{\mu}_{\nu} \overline{\varphi}_{\lambda} + \delta^{\mu}_{\lambda} \overline{\varphi}_{\nu} - g_{\nu \lambda}\,\overline{\varphi}^{\, \mu}) \, \label{eq conformal equivalence}
\eeq 
for  some differential form  $\overline{\varphi}$. It is important to realize that this differential form need {\em not be exact} (locally integrable), as one would expect if the characterization is read  in the context of Riemannian geometry and their Levi-Civita connections.\footnote{\citet[833]{Matveev/Trautman} makes this restriction, see sec. \ref{subsec: Matveev/Trauman}. }

On this basis  Weyl easily proved the following
\begin{theorem}[Weyl 1921]
 A Weylian metric is  uniquely determined by its  projective and its conformal structures.\footnote{``Satz 1. {\em  Projektive und konforme Beschaffenheit eines metrischen Raums bestimmen dessen Metrik eindeutig}'' \citep[p. 196]{Weyl:projektiv_konform}.}
  \end{theorem}
The theorem deals with the comparison of two different classes $[\Gamma]_p,\, [\Gamma]_c$ of affine connections with non-empty intersection (because both arise by abstraction from the same presupposed Weylian metric). 
If $\Gamma$ and $ \Gamma'$ are now any two in  $[\Gamma]_p \cap [\Gamma]_c $, both arising as affine connections of  Weylian metrics $[(g, \varphi)]$, $[(\tilde{g}, \tilde{\varphi})]$, it has to be shown that the latter are identical. In any case the 
 Riemannian components of such two Weylian metrics are conformally equivalent, and the $\Gamma, \, \Gamma'$  are related by (\ref{eq conformal equivalence}).  Both Weylian metrics can  be gauged to identical  Riemannian components, i.e.  be given in gauges of the form $(g,\varphi)$,  resp. $(g,\varphi')$. The differential form $\overline{{\varphi}}$ in (\ref{eq conformal equivalence}) is then  $\overline{{\varphi}} = \varphi-\varphi'$. As $\Gamma$ and $\Gamma'$ are also projectively equivalent, their difference $\Delta = \Gamma-\Gamma'$ changes a vector $\xi$ such that $\Delta^{\mu}_{\nu \lambda}\, \xi^{\nu}\xi^{\lambda} \sim \xi^{\mu} $ (here $\sim$ stands  for proportionality). Plugging in (\ref{eq conformal equivalence}) for $\Delta$ shows that $g_{\nu \lambda}\xi^{\nu} \xi^{\lambda} \, \overline{{\varphi}}^{\mu}$ is proportional to $\xi^{\mu}$ for  all vectors $\xi$. This implies $\overline{{\varphi}}=0$.

Weyl emphasized the physical importance of this theorem: The conformal structure characterizes the {\em causality} relations in spacetime. The projective properties of space are an expression of, in later terminology, the {\em gravito-inertial} structure of spacetime. Weyl described it in the following way:
\begin{quote}
\ldots the tendency of persistence of the direction for a moving material particle, which impresses a certain `natural' motion on it, once it has been set free in a specified world-direction, 
is  the very unity of inertia and gravity, which Einstein put in the place of both, although   a suggestive name for it is still lacking \citep[p.196]{Weyl:projektiv_konform}.\footnote{``\ldots die Beharrungstendenz der Weltrichtung eines sich bewegenden materiellen Teilchens, welche ihm, wenn es in bestimmter Weltrichtung losgelassen ist, eine bestimmte `nat\"urliche' Bewegung aufn\"otigt, ist jene Einheit von Tr\"agheit und Gravitation, welche Einstein an Stelle beider setzte, f\"ur die es aber bislang an einem suggestiven Namen mangelt'' \citep[p.196]{Weyl:projektiv_konform}.}
\end{quote}

At the time of the 5th edition of RZM Weyl could claim to have a well-rounded generalized concept of  infinitesimal gauge geometry. This geometry was mathematically well developed (uniquely determined affine connection, curvature properties), had a convincing conceptual underpinning in terms of basic physical structures (causality, inertio-gravitational persistence), and could even be given a transcendental philosophical backing  (PoS). 

On the other hand, Weyl's early enthusiasm for the capacity of his theory for leading to a  unified theory of fields and matter had become cracks because of internal technical difficulties of the program (extremely complicated differential equations).  Even stronger doubts arose from the growing impression of physicists in the environment of Sommerfeld that quantum physics might make the attempts for a classical field theory of matter obsolete anyhow. Weyl was well aware of such doubts  and tended to share them.

\subsection{Withdrawel of scale gauge by Weyl after 1927/29}
\label{subsec: Withdrawel} 
Already  
in the early 1920s  E. Schr\"odinger had the idea that  Weyl's gauge principle might be useful  in a modified form for dealing with the phase of complex wave functions in quantum mechanics, rather than for scale in gravity. This idea was taken up  and elaborated by  F. London  and independently by V. Fock after the turn towards the ``new'' quantum mechanics.  Weyl endorsed it in the first edition of his book on {\em Gruppentheorie und Quantenmechanik} \citep[1st ed., p. 87]{Weyl:GQM}. A year later Weyl  extended this to a general relativistic approach to spinor fields. This led to the now well known representation of the electromagnetic potential by a connection with values in the Liealgebra of the phase group $U(1)$.\footnote{See \citep[p. 274ff.]{Vizgin:UFT}, \citep{Straumann:DMV,Scholz:Fock_Weyl,Afriat:2013} and N. Straumann's contribution to this volume.}
At the end of the decade Weyl was quite fond of this migration of the gauge idea from scale to phase and considered it as a definitive answer  to the question how electromagnetism ought to  be understood as a gauge theory.

The 1930  Rouse Ball lecture at  Cambridge university gave him an opportunity  for  explaining his view of the  program of geometrical unification to a wider scientific audience. He explained  his own theory of 1918 and summarized its critical reception by physicists. He  reviewed Eddington's approach to unification by affine connections and Einstein's later suppport for that program, always in comparison with his own   ``metrical'' unification of 1918. He concluded that in hindsight one could consider both theory types merely as  ``geometrical dressings (geometrische Einkleidungen) rather than proper geometrical theories of electricity''. He  added with irony that the struggle between the metrical and affine unified field theories (UFT), i.e. his own 1918 theory versus Eddington/Einstein's,  had lost  importance. In 1930  it could no longer be the question   which of the theories would ``prevail in life'', but only ``whether the two twin brothers had  to be buried in the same grave or in two different graves''  \citep[343]{Weyl:Rouse_Ball}.  
All in all Weyl perceived a scientific devaluation of the UFT's of the  1920s,  resulting from developments in the second part of the decade:
\begin{quote}
In my opinion the whole situation has changed during the last 4 or 5 years  by the detection of the matter field. All these geometrical leaps (geometrische Luftspr\"unge) have been premature, we now return to the solid ground of physical facts. \cite[343]{Weyl:Rouse_Ball}
\end{quote}

He continued to sketch the theory of spinor fields,  their phase  gauge and its  inclusion into the framework of general relativity along the lines of his 1929 articles. Weyl emphasized  that,  in contrast to the principles on which the  classical unified field theories had been built, 
the new principle of phase gauge ``has grown from experience and resumes a huge treasury of experimental facts from spectroscopy'' (ibid. 344). He still longed for 
safety,  just as much  as at the time after the First World War, when he had designed his first gauge unification. Now he no longer expected to achieve it by geometric speculation,  but tried to anchor it in  more solid grounds: 
\begin{quote}
By the new gauge invariance the {\em electromagnetic field now becomes a necessary appendix of the matter field, as it had been attached to gravitation in the old theory. } \cite[345, emphasis in original]{Weyl:Rouse_Ball}
\end{quote} 
In this way Weyl made it  clear that he had changed his perspective. He no longer saw a chance in  attempts to derive matter in highly speculative approaches from mathematical structures devised to geometrize force fields; he now set out  searching  for  mathematical representations of matter which was based on the  ``huge treasury'' of experimental knowledge. For him, this was reason enough to prefer  the view that the electrical field ''follows the ship of matter as a wake, rather than gravitation'' (ibid.). 

This paper indicates a  re-evaluation  of Weyl's view of geometry with regard to those of  the early 1920s. This change of mind  took place at the turn to the 1930s and is not yet present in the first German edition  of his book {\em Philosophie der Mathematik und Naturwissenschaften}  \citep{Weyl:PMN}, but it is in the English translation for which the author formulated text amendments and changes during  1948/49 \citep{Weyl:PMNEnglish}. These changes were taken over  into the German  third edition (after Weyl's death).

In a talk  with the title   {\em Similarity and congruence} given during the  time when he worked on the changes for {\em Philosophy of Mathematics and Natural Science}  Weyl discussed the topic of {\em automorphism groups} as a  clue for establishing objectivity of  symbolic knowledge  in mathematics and in physics.\footnote{This talk is published  in \citep{Weyl:similarity}} Weyl hoped to be able to clearly  distinguish between {\em physical} and {\em mathematical automorphisms}. The latter were characterized by him  as the  normalizer of the former -- leaving open the question in which larger group the normalizer was to be taken. For classical physics this was relatively simple: the physical automorphisms of classical physics are given by the Galilei group (including Euclidean congruences), and the mathematical automorphisms are the similarity transformations extending the Galilei group. 
Weyl therefore used the pair similarities/congruences also in the general case as synonymous with the dichotomy mathematical/physical automorphisms.

Of course it is a difficult question to decide what the physical automorphisms are; but Weyl was sure that this is a central  task of physics:
\begin{quote}
The physicist will question Nature to reveal him her true group of automorphisms \citep[p. 156]{Weyl:similarity}.
\end{quote}
For relativistic physics, so Weyl argued in 1949,  the physical automorphisms were given by the diffeomorphisms  $\textsl{Diff}\,(M)$ of the spacetime manifold $M$, extended by point dependent operations of   $G=SO(1,3)\times U(1)$. In hindsight Weyl's physical automorphisms of relativistic physics may be  considered as an informal characterization of a  gauge group $\mathfrak{G}(P)$ {\em ante letteram} with respect to a principal bundle $\mathfrak{P}$ over $M$ with the group $G$. In his view the mathematical automorphisms were then the corresponding ``similarities'' given by the extension from $G$ to $\widetilde{G}= G \times \R^+$.\footnote{For more details see \citep{Scholz:2018Weyls_search}.} 
In the view of the mature Weyl, the contemporary knowledge of the 1930/40s in quantum physics clearly  spoke for reducing the physical automorphisms to the group  $G$. 
\begin{quote}
The atomic constants of charge and mass of the electron atomic
constants and Planck's quantum of action $\hbar$, which enter the
universal field laws of nature, fix an absolute standard of length,
that through the wave lengths of spectral lines is made available
for practical measurements \citep[p. 161]{Weyl:similarity}.\footnote{Similarly in PMN.}
\end{quote}

In the 1940s Weyl  no longer considered the local scale extension from $G$ to $\widetilde{G}$ as part of physics, but of mathematics only.
Taking up his language of 1919 in the discussion with Einstein,  the  laws of quantum mechanics and the universal constants $(\hbar, e, m_e)$ had now taken over the role of the {\em central office of standards}. For Weyl, this was  a definitive good bye  to the idea of localized  standards of length/time due to an adaptation of atomic oscillators to local field constellations and from the view that his scale gauge geometry is an adequate conceptual framework for gravity and field theory. 

He did not care about an interesting observation made by {\em Jan A.  Schouten} and {\em Jan Haantjes} in the 1930s,  who argued that not only the (vacuum) Maxwell equations, but also the equations of motion of test mass particles  (with or without Lorentz forces) can be written in a scale covariant form, if only the mass parameter $m$ is being transformed scale covariantly with quadratic Weyl weight $-\frac{1}{2}$ \citep{Schouten/Haantjes:1934,Schouten/Haantjes:1936,Haantjes:1941}. At the time, this was  an unusual point of view. Pauli argued that only the massless Dirac equation could  be considered as scale invariant \citep{Pauli:1940}; a similar point of view had been expressed by Weyl in the context of the general relativistic Dirac equation \citep{Weyl:1929Dirac}. Schouten's and Haantjes' common proposal to consider mass parameters as scaling quantities indicated a path toward including massive spinor fields into a basically conformal -- or correspondingly a Weyl geometric -- framework; but at the time it was not widely noted. 

\section{A new start for Weyl geometric gravity in the 1970s}
\label{sec: Retake}

\subsection{New interests in local scale and conformal transformations}
\label{subsec: new interests}
In the early 1950s and 1960s {\em Pascual Jordan},  later {\em Robert Dicke} and {\em Carl Brans} (JBD) proposed a widely discussed   modification of Einstein gravity  \citep{Jordan:Schwerkraft,Brans/Dicke,Dicke:1962}. 
 Essential for their approach was a (real valued) scalar field $\chi $, coupled to the  Hilbert action term. Its Lagrangian density was
 \beq \mathcal{L_{\textsl{JBD}}} = (\chi  R - \frac{\omega }{\chi }\partial ^{\mu}\chi \, \partial _{\mu}\chi  ) \sqrt{ | det \,  g|}  \; , \label{L_JBD}
\eeq 
where $\omega $ is a free parameter of the theory and $R$  the scalar curvature of the Riemannian metric $g$. 
Jordan had started from a projective version of a 5-dimensional Kaluza-Klein approach. He arrived at a Lagrangian of form (\ref{L_JBD}) only after several steps of simplifications and interpreted the scalar field as a varying gravitational parameter (replacing the gravitational constant).\footnote{For technical details see \citep[pp. 31ff.]{Goenner:2012}, for the physical interpretation and the historical context \citep[pp. 45ff, 65ff.]{Kragh:VaryingGrav}; a conceptual analysis will be given in \citep{Lehmkuhl:Jordan}. }
 For $\omega \rightarrow \infty $ the theory has Einstein gravity as limiting case. 
All three authors allowed for conformal transformations of the metric, $\tilde{g} = \lambda g$; but only Brans and Dicke understood them as an expression for a local scale transformation  under which also the scalar field $\chi $ transforms with (quadratic length) weight $-1$ (matter fields and energy tensors $T$ with weight $w(T)=-\frac{1}{2}$ etc.).

Jordan started considering conformal transformations  only  after Pauli had made him aware of such a possibility; he discussed them in the second edition of his book \citep{Jordan:Schwerkraft}.  Pauli must have been  aware of the closeness of this principle to  Weyl's scale geometry; in his youth he had been one of the  experts for it. But neither he nor Jordan, looked at the new scalar tensor theory  from this point of view. The migration of the gauge idea from scale to phase geometry seems to have been considered  by them, like by Weyl himself, as definitive. The US-American authors nearly a decade later  were probably not even aware of the parallel.\footnote{In retrospect C. Brans wrote regarding this question:``I believe (but am not sure) that I knew of other UFT's,
especially Kaluza-Klein, but do not know if I was
aware of Weyl's conformal work.  I wish I could be
more definite, but the best answer I can give to
your question of whether Bob or I was aware of Weyl's
conformal scalar field is `probably not'.'' (e-mail of C. Brans to the author, 19 June, 2012) }

Maybe this ignorance was  an advantage. Dicke did not hesitate when advocating  conformal rescaling. He frankly declared it as an obvious  postulate  that the  ``{\em  laws of physics must be invariant under a general coordinate dependent  change of units}'' (see quote  in the introduction,  p. \pageref{Dicke-evident}).
He did not mention that this would demand a basic restructuring, or at least reformulation of fundamental theories, although he  must have been aware of it.
The postulate itself agreed with Weyl's intentions of 1918. In a long appendix to his letter of 16 Nov. 1918  to Einstein,  published in extended form during the next year, Weyl had stated
\begin{quote}
 Einstein's present theory of relativity (\ldots)  only deals with the arbitrariness of the coordinate system; but it is important to gain a comparable foundational stance with regard to (\ldots) the arbitrariness of the measurement of units \citep[p. 55]{Weyl:Erweiterung}.\footnote{``Die bisherige Einsteinsche Relativit\"atstheorie bezieht sich nur auf (\ldots) die Willk\"urlichkeit des Koordinatensystems; doch gilt es eine ebenso prinzipielle Stellungnahme zu (\ldots) der Willk\"urlichkeit der Ma{\ss}einheiten zu gewinnen'' \citep[p. 55]{Weyl:Erweiterung}. }  
\end{quote}
Knowing more about  Weyl's trajectory, who had recanted this viewpoint in the 1940s,  might only have been a hindrance for developing  the new scalar tensor theory, or at least the inclusion of the conformal transformations viewpoint into it. 

By ``coordinate dependent change of units'' Dicke and Brans  indicated a point dependent rescaling of basic units. In the light of 
 the relations  established by the fundamental constants (velocity of light $c$, Planck constant $\hbar$, elementary charge $e$ and Boltzmann constant $k$) all units can be expressed in terms of one independent fundamental unit, e.g. time, and the fundamental constants (which, in principle can be given any constant numerical value, which then fixes the system).\footnote{For the recent revision of the  international standard system SI see \citep{Hehl/Laemmerzahl:SI}. It has implemented  measurement definitions with time as only fundamental unit, $u_T = 1\, s$ such that ``the ground state hyperfine splitting frequency of the caesium 133 atom  $\Delta\nu(^{133}Cs )_{\mbox{hfs}}$ is exactly $9\,192\,631\,770$ hertz''  \cite[24f.]{SI:2011}. In the {\em New SI}, four of the SI base units, namely the kilogram, the ampere, the kelvin and the mole, will be redefined in terms of invariants of nature (www.bipm.org/en/si/new$_{-}$si/). The redefinition of the meter in terms of the basic time unit by means of the fundamental constant $c$ was implemented already in 1983. Point dependence of the time unit because of the locally varying gravitational potential on the surface of the earth is inbuilt in this system. For practical purposes it can be outlevelled by reference to the {\em SI  second on the geoid} (standardized by the International Earth Rotation and Reference Systems Service IERS). Some, only  seemingly paradoxical, consequences in the description of astronomical distances are being discussed by T. Sch\"ucker, this volume. \label{fn SI} } 
 Thus only one essential scaling degree of units remains, and 
Dicke's  principle of an arbitrary point dependent unit choice comes down to  a  ``passive'' formulation of Weyl's  localized similarities  in his scale gauge geometry, where dimensional constants are to be treated as scale covariant scalar fields with the respective Weyl weights. 

A closer look shows that  Dicke's boastful postulate that the ``laws of physics
must be invariant'' under point dependent rescaling  was not fully realized in JBD theory. The {\em modified Hilbert term} of (\ref{L_JBD}) is formulated in terms of the  Riemannian scalar curvature  and {\em is not  scale invariant}. The practitioners of JBD theory understand it as defined in a specific scale (the  Jordan frame) and apply  well known correction terms under conformal rescaling. This defect can easily be cured if one reformulates the theory in terms of a simple version, i.e.  integrable,  Weyl geometry (see sec. \ref{subsec: Philosophical reflections}).

In spite of such a (minor) formal deficiency,  the  three proponents of JBD theory unknowingly brought their approach quite close to  Weyl geometry by fixing a {\em unique affine connection} rather than changing it with the conformal transformation. They postulated the 
 Levi-Civita connection $ \Gamma := \Gamma(g) $ of the Riemannian metric $g$ in
 (\ref{L_JBD}), called the {\em Jordan frame}, as  unchanging  under conformal transformations.  In a different scale gauge, or {\em frame},  $\tilde{g}= \lambda g$  they just had to express $\Gamma$ in terms of the Levi-Civita connection $\Gamma(\tilde{g})$ plus correction terms in derivatives of $\lambda$ (equivalent to (\ref{eq Gamma})). Let us summarily denote these additional terms by  by $\Delta (\partial  \lambda)$,
 then
\[ \Gamma = \, _{\tilde{g}}\hspace{-0.05em}\Gamma  + \Delta (\partial  \Omega) \; . \] 
{\em Roger Penrose} noticed that the  additional terms of the (Riemannian) scalar curvature are exactly cancelled by the partial derivative terms of the kinematical term of $\chi$ if and only if $\omega = - \frac{3}{2}$. In this case, and in JBD theory only in this case, the Lagrangian (\ref{L_JBD}) is conformally invariant \citep{Penrose:1965}.

 Probably the protagonists considered the invariance of the affine connection $\Gamma$ as a consequence of the principle that  the ``laws of nature'' have to be considered as  invariant under conformal rescaling. If the  trajectories of test bodies are governed by the  gravito-inertial ``laws of physics'' they should not be subject to change under a transformation of units. The same must then hold for the affine connection  which can be considered a mathematical concentrate of these laws.  Thus passive {\em conformal rescaling}, in addition to  {\em fixing an affine connection}, have become {\em basic tools of  JBD theory}.\footnote{For an outline of conformal transformation as used in JBD see \citep[app. D]{Wald:GR}; surveys on the actual state of JBD theory and its applications to cosmology are given in \citep{Fujii/Maeda,Faraoni:2004}.}

JBD theory originated at  a time when symmetry aspects attracted more and more attention by high energy particle physics. Here symmetries were often not considered as  exactly realized, but as somehow ``hidden'' or even ``broken''.\footnote{See \citep{Borrelli:2015,Borrelli:isospin}.}
Also  conformal transformations were now being reconsidered  in field physics. {\em Werner Heisenberg} took up the idea of Haantjes and Schouten and proposed to consider rescaling of mass parameters as a legitimate symbolic procedure \citep{Heisenberg:1957};  and so did other elementary particle physicists.\footnote{Among them J. \citet{Wess:1959,Wess:1960} and H.    \cite{Kastrup:1962,Kastrup:1962AnnPh}.}
 In this context it became a standard procedure to associate scale weights to physical fields.  
 For a detailed historical report see \citep{Kastrup:2008}.
 
 Of course, elementary particle physicists were not interested in conformal transformations of general Lorentzian manifolds; they  dealt exclusively with  the  group $\textsl{Conf}\,(\widetilde{\M})$ of 
 conformal transformations of the (conformal) compactification $\widetilde{\M}$, of Minkowski space. The group can be expressed as the subgroup of projective transformations in real projective space $P^5(\R)$ with projective coordinates $[u_0, \ldots u_5]$,  which leaves the hyperbolic quadric $ Q(2,4)$ 
 \[ \qquad u_0^2 + u_5^2 - \sum_{j=1}^4 u_j^2 = 0
 \]
  invariant. The embedding of Minkowski space $\M \hookrightarrow Q(2,4) \subset P^5(\R)$ is given by 
 \[ (x_0, \ldots x_3) \longmapsto [u_0, \ldots, u_3, \frac{1}{2}(|x|^2+1), \frac{1}{2}(|x|^2 - 1)] \, , \]
 with $|x|^2 = x_o^2 - \sum_{j=1}^3 x_j^2 $.
 This was well known at the time.\footnote{Such a  representation of $\textsl{Conf}\,(\widetilde{\M})$ is sketched, e.g.,  in \citep[p. 302f.]{Weyl:RZM5}.} 
But  the understanding of  the  reciprocal transformations (inversions) 
with regard to Lorentzian  hyperboloids rather than to euclidean spheres was hampered for some time because of confused attempts to interpret them as an expression for transformations of relatively accelerated observer systems. Such proposals  had been made in the 1930/40s, among others by \citet{Haantjes:1941}. They  lived until the mid 1960s, although it was clear to a part of the community that this was misleading.\footnote{\citep[p. 659ff.]{Kastrup:2008}}
{\em Hans Kastrup} for example,  developed an interpretation with point dependent measuring standards for describing the effects of hyperboloidal inversions and studied the   representations of the full conformal group on scalar and spinor fields with the hope that the inversions would, at least, turn out to be approximative symmetries of a conformal field theory \citep{Kastrup:1966}, . In the discussion of the point dependent standards he explicitly referred back to Weyl's scale gauge method of 1918 \citep[p. 150]{Kastrup:1966}, although the wider scope of Weyl geometry was of no importance for his investigations.
 
The questions regarding the inversions posed in  the GR community were  different.  Penrose gave a detailed geometrical analysis of the light cone at infinity in Minkowski space \citep{Penrose:ConfInf}. From such a point of view it became  clear that  the    reciprocal transformations are mappings between conformal infinity and finite light cones, which induce a conformally deformed metric (with regard to the Minkowski metric) in the neighbourhood of the latter. In general relativity such transformations were useful for studying fields in asymptotic flat spacetimes. They even became a step towards better understanding the  asymptotic behaviour of general relativistic spacetimes also  in  more general cases.\footnote{\citep{Frauendiener:ConfInf}.} In this sense, hyperboloid inversions, or their relatives in more general spacetimes, remained at best  ``mathematical automorphisms'' in the language of Weyl, useful for technical reasons  rather than for physical ones.
 
In the  high energy context, on the other hand, reciprocal transformations had a different appeal, in particular if applied to the energy-momentum space (in a physical sense dual to Minkowski spacetime). If everything went well, they  seemed to allow conformal field theory searching for   relationships between field states in  extremely high energies/momenta and extremely small energy/momentum states close to a finite light cone, even though such inversion symmetries might perhaps  be ``hidden'' or ``broken'' and indicate the relations only in a modified form. In this sense, the  understanding of conformal symmetries in high energy physics was characterized by an active understanding in the perspective of extending the ``physical'' automorphisms in Weyl's description of 1948/49. 
 
Such differences of outlook my have contributed to the unhappy constellation that the interchange between general relativity and conformal field theory (in Minkowski space) remained rather weak. As far as I can see, there were   mutual   methodological challenges between the fields, but no  a closer exchange.

\subsection{Weyl geometric gravity with a scalar field and dynamical scale connection: Omote, Utiyama, Dirac}
\label{subsec: WST}
In 1971   {\em Minoru Omote}, Tokyo, introduced a scale covariant scalar field $\phi$ coupling to the Hilbert term and scaling like in  JBD theory into the framework of Weyl geometry \citep{Omote:1971}.\footnote{This was more than a year before the Trieste symposium at which Dirac talked about the same question (see below), while it seems that Omote's  paper remained unknown to him.} A second paper by Omote followed after the publication of a paper by P.A.M. Dirac with a similar proposal  \citep{Dirac:1973} and after R. Utiyama  had jumped in \citep{Omote:1974} (for both authors see below).
{\em Alexander Bregman},  
 at that time working at Kyoto, was inspired by Omote's proposal  to separate  localized rescaling from Weyl's geometrical interpretation of the infinitesimal length transport  \citep{Bregman:1973}. He argued that the point-dependent scale transformations  could be treated ``analogous to the introduction of a space-time dependence into the constant parameters of Isospin or Poincar\'e transformations'' (ibid. p. 668). This brought the approach closer to what high energy physicists were doing at the time, although only the scale extended Poincar\'e group was localized, not the complete conformal group of Minkowski space. The global scale dimensions  $d$ of a physical field $X$ could then be  taken over as ``Weyl weight'' of $X$  to the localized theory   \citep[p. 668]{Bregman:1973}. With  such a proposal he followed the lines of the  research program for constructing general relativistic theories of gravity by ``localizing'' the symmetries of the Poincar\'e group, which had been opened by D. Sciama and T. Kibble \citep{Kibble:1961,Sciama:1962}.\footnote{See \citep{Blagojevic/Hehl}.}

 A little later,  more or less parallel to 
 Dirac,  {\em Ryoyu Utiyama}, Toyonaka/Osaka,  joined the new start of a Weyl geometric approach to gravity \citep{Utiyama:1973}. Bregman's approach seemed to have triggered his interest which lay mainly in  elementary particle physics.\footnote{In his first paper of 1973 Utiyama did not mention Omote. This changed in later papers, see  the references of \citep{Utiyama:1975a}.}
 Different from Dirac and, of course also from Weyl, he saw in  the nontrivial  Weylian scale connection $\varphi$ a candidate for a new fundamental field of high energy physics, the  Weyl gauge field or, later, the {\em Weyl boson}.  In a series of papers \citep{Utiyama:1973,Utiyama:1975a,Utiyama:1975II}  he   ventured toward  a bosonic interpretation  of $\varphi$ and   presented his results at the Seventh International Conference on Gravitation and Relativity (Tel Aviv, June 1974). Utiyama emphasized that in Weyl geometry a scalar field $\phi$ of weight $-1$ could serve as a  kind of {\em measure field}  (Utiyama's terminology). \label{Utiyama}
 With respect to it  gauge invariant measurable quantities could be defined for physical observable without assuming a breaking of the scale gauge symmetry  \citep{Utiyama:1973,Utiyama:1975II}.

 Utiyama proposed to explore the ordinary Yang-Mills Lagrangian term for a Weylian scale connection
\beq   \mathcal{L}_{\varphi}= - \varepsilon  \frac{1}{4}  f_{\mu \nu } f^{\mu \nu } \sqrt{|det\, g|} \quad (\mbox{here with}\; \varepsilon =1)  \label{eq L_varphi Utiyama}
\eeq
\citep[eq. (2.4)]{Utiyama:1975II}.\footnote{Dirac included a similar scale curvature term in his Lagrangian, but with another interpretation (see below).} 
He studied conditions under which ``Weyl's gauge field''  admitted plane wave solutions, and came to the conclusion that they would be  tachyonic, allowing  superluminal propagation of perturbations.  In Utiyama's  view the   ``boson'' had therefore to be confined to the interior of matter particles. Nevertheless he thought that this ``unusual field $\varphi_{\mu }$ might play some role in establishing a model of a stable elementary particle'' \citep[2089]{Utiyama:1973}.  
This view was not  accepted by all his readers. {\em  Kenji Hayashi} and {\em Taichiro Kugo}, two younger colleagues from Tokyo resp. Kyoto, reanalysed  Utiyama's calculations and argued that, with slight adaptations of the  parameters, the sign $\varepsilon $ in (\ref{eq L_varphi Utiyama}) could  be switched. Then an ordinary, at least non-tachyonic, field would result \citep[340f.]{Hayashi/Kugo:1979}.

Even then the scale connection would still have strange physical properties. After   introducing  Weyl geometric spinor fields and their Lagrangians in terms of   scale covariant derivatives, the two physicists showed that the  scale connection terms cancel in the spinor action. In their approach neither the scalar $\phi$-field nor the scale connection $\varphi$  coupled directly to spinor field or to the electromagnetic field. The new fields $\phi,\, \varphi$ seemed to characterize an extension of the gravitational sector with no direct interaction with the known elementary 
particles.  
At the very moment that a Weylian scale connection $\varphi$ was  interpreted as a physical field beyond electromagnetism,  it started to puzzle its investigators. It seemed to pose more riddles than it was able to solve. It  did not to  couple  to matter fields  (Hayashi/Kugo),  looked either  tachyonic (Utiyama) or, as we shall see below (Smolin, Nieh, Hung Cheng), appeared to be of Planck mass, far beyond anything observable. \\[-0.5em]


Independent of the Japanese physicists, and mor or less at the same time,   {\em Paul Adrien Maurice Dirac}  brought Weyl geometry back into the  rising field scalar-tensor theories, although with different interpretations from his Japanese colleagues in mind \citep{Dirac:1973,Dirac:LNH}. 
His motivation had two components which may look strange from today's point of view. He started from the speculation that some long noticed  interrelations of certain large numbers in physics indicated a deep structural law of the universe (the ``large number hypothesis'') and the idea of a varying gravity. Both ideas had their origin in the 1930s,  and also P. Jordan had taken them up.\footnote{For  Dirac's and Jordan's  ideas on   the large number hypothesis, varying gravity and a surprising link to geophysics  in  the 1930s  see the detailed study in  \citep{Kragh:VaryingGrav}.}
Dirac presented his ideas at the occasion of a  symposium  at Trieste  1972, honouring  his 70th birthday.  The talk remained unpublished but participants report that its content was close to publication in the following year \citep{Dirac:1973}.\footnote{\citep[p. 249 footnote]{Charap/Tait}} 

Dirac's paper was important for the transmission of knowledge between generations. He introduced his readers to Weyl geometry which was no longer generally known among  younger  physicists, 
following Eddington's notation and terminology of ``co-invariants'' for scale covariant fields \citep{Eddington:Relativity}. He also  added the important concept of the {\em scale-covariant derivative } $D$ , respectively  $D_{\mu}$ (in later terminology and notation),  for scale covariant fields  to the methodological arsenal of Weyl geometry (see below, eq. (\ref{scale covariant D})). It is  a necessary modification of the  covariant derivative of scale covariant fields for  arriving again at a scale covariant fields. He called it, virtually stuttering, the ``co-covariant'' derivative. \label{Dirac co-co}

Similar to Jordan/Brans/Dicke  -- and like Omote/Bregman/Utiyama, the publications of  which he apparently did not know -- Dirac introduced a scalar field called $\beta$ and rescaling with the (non-quadratic) length  weight $-1$ and coupled it to the Hilbert term   expressed by  the sign-inverted Weyl geometric scalar curvature $R$.\footnote{The qualifications ``sign inverted'' refers to the  sign  convention which agrees with the definition $\textsl{Riem}\,(Y,Z)\,X=\nabla_Y \nabla_Z X - \nabla_Z \nabla_Y X - \nabla_{[Y,Z]}X$, i.e. 
$R^{\mu}_{\;\; \nu \lambda \kappa}=  \partial_{\lambda} \Gamma^{\mu}_{\nu \kappa}  - \partial_{\kappa} \Gamma^{\mu}_{\nu \lambda}  +  \Gamma^{\alpha}_{\nu \kappa} \Gamma^{\mu}_{\alpha \lambda}  - \Gamma^{\alpha}_{\nu \lambda} \Gamma^{\mu}_{\alpha \kappa}  $. It is  preferred in the mathematical literature including \citep[5th ed., 131]{Weyl:RZM} and also used in the majority of recent physics books. 
  \label{fn curvature conventions}} 
  Thus he replaced Weyl's  gravity  {\em Ansatz} in the Lagrangian, using square curvature terms,  by a scale invariant  Lagrangian  ($\mathcal{L}_{Dir} =  {L}_{Dir} \sqrt{|g|}$) of first order in $R$,   
\beq   L_{Dir} =-  \beta^2 R + k D^{\lambda}\beta\, D_{\lambda}\beta + c \beta^4 + \frac{1}{4} f_{\mu \nu}  f^{\mu \nu} \; .     \label{scale invariant Dirac action}  \eeq
The quartic potential for $\beta$ was demanded by  scale invariance.  $f=d\varphi$ denoted the Weylian scale curvature. Different from the Japanese physicists, Dirac stuck to the outdated interpretation of the scale connection $\varphi$ as the   potential of the electromagnetic (Maxwell) field $F_{\mu \nu}$
\beq  F_{\mu \nu} = f_{\mu \nu} \; . \eeq 
In the sequel I  call this the {\em electromagnetic    (em) dogma} .

For the coupling constant of the kinetic term,  $k=6$, the contributions of the scale connection to  the Lagrangian cancel essentially.\footnote{They reduce to boundary terms and thus are  variationally negligible.}
Dirac knew that for $k\neq 6$ a large mass term for the Weyl field $\varphi$ arises, which would destroy the    {\em em}-dogma interpretation; so he chose $k=6$.  
As a result Dirac could write   the  Lagrangian   in a form using only the (sign inverted) Riemannian  component $_g\hspace{-0.1em}{R} $  of the scalar curvature
\beq       \mathcal{L}_{Dir\, 1} =  -\beta^2\,  _g\hspace{-0.1em} {R} + 6 \partial ^{\lambda}\beta\, \partial _{\lambda}\beta + c \beta^4 + \frac{1}{4} f_{\mu \nu}  f^{\mu \nu} \; ,                    \label{Dirac action} \eeq 
It was already known to be conformally  invariant \citep{Penrose:1965}, and only the specific Weyl geometric interpretation of the {\em em} potential was added. 

If it had not been  Dirac, such an approach would probably not have   attracted much interest in the physics community. But he also 
 derived dynamical equations and the Noether identities for diffeomorphisms and scale transformations.
For a vanishing {\em em} field, $f_{\mu \nu}=0$, he distinguished the {\em Riemann gauge} with $\varphi =0$ (called by him ``natural gauge'') from the {\em Einstein gauge}  (with the gravitational parameter constant, $\beta =1$) and a hypothetical  {\em atomic gauge} characterized as  ``the metric gauge that is measured by atomic apparatus'' (Weyl's ``natural gauge'')  and  warned that 
``all three gauges are liable to be different''  \citep[411]{Dirac:1973}. 

At the end of his article Dirac discussed why one should believe in the proposed ``drastic revision of our ideas of space and time''. He announced another part of his research agenda, which was {\em independent} of the large number hypothesis:
\begin{quote}
There is one strong reason in support of the theory. It appears as one of the fundamental principles of Nature that the equations expressing basic laws should be invariant  under the widest possible group of transformations \ldots  The passage to Weyl's geometry is a further step in the direction of widening the group of transformations underlying physical laws. 
 \citep[418]{Dirac:1973}
\end{quote}

So far, Dirac's explanations were  close to the view of  Brans and Dicke.  He  followed a tendency of the time  for probing possible extensions of  the symmetries (automorphisms) of fundamental physics and saw a new chance for Weyl geometry to play its part in such an endeavour.  

 Dirac's proposal for reconsidering Weyl geometry in a modified theory of gravity was taken up by field theorists, gravitational physicists and a few astronomers.
An immediate and often quoted paper by {\em Vittorio Canuto} and coauthors gave a broader and more detailed introduction to Dirac's view of Weyl geometry in gravity and field theory \citep{Canuto_ea}. The opening remark of the paper   motivated  the renewed interest in Weyl geometry with  actual developments in high energy physics:
\begin{quote}
In recent years, owing to the scaling behavior exhibited in high-energy particle scattering experiments there has been considerable interest in manifestly scale-invariant theories.
\citep[1643]{Canuto_ea}
\end{quote}
With the remark on ``considerable interest in manifestly scale-invariant theories'' in high energy physics the authors  referred  to  scaling in high energy physics  and, in particular, the seminal paper  \citep{Callan/Coleman/Jackiw}.
But the authors were careful not to claim field theoretic reality for Dirac's scalar function  $\beta$  \citep[1645]{Canuto_ea}. They rather  developed model consequences for the approach in several directions:  cosmology, including the ``large number hypothesis as a gauge condition'' (ibid. 1651), modification of the Schwarzschild solution in the Dirac framework, consequences for planetary motion, and stellar structure. At the end the authors indicated certain heuristic links to gauge fields in high energy physics of the late 1970s.

 Dirac's retake of Weyl geometric gravity and field theory had wider repercussions than might be expected if one takes his insistence on the {\em em}-dogma into account. The scope of the approach was much wider without it, and a lot of authors with different backgrounds and  diverging research interests started again to explore the potential of Weyl's geometry for widening the geometric framework of gravity, most of them without subscribing to  the {\em em}-dogma. Here  I can only hint summarily at a sample from different groups or individual authors who took part in this exploration during the following decades until, roughly, the end of the century. Among them were  P. Bouvier and A. Maeder, Geneva, from {\em astronomy} \citep{Bouvier/Maeder:WeylGeometry,Maeder:lv_planets,Maeder:1980};\footnote{In recent publications A. Maeder has taken again up this line of research \citep{Maeder:2017dm,Maeder:2017CMB,Maeder:2018}, see section \ref{subsec: cosmology and dm}.} 
  N. Rosen, the former collaborator of Einstein, and his PhD student M. Israelit came  from {\em gravitational theory} \citep{Rosen:1982,Israelit/Rosen:1992dm,Israelit/Rosen:1995dm,Israelit:1999Book},  and from {\em cosmology} M. Novello with a growing group of researchers in Brazil 
 \citep{Novello/Oliveira:1986,Novello/Oliveira_ea:1993}. Other authors like L. Smolin, W. Drechsler, his PhD student H. Tann  and H. Cheng, working in {\em  fundamental} or {\em high energy physics},  also took up Weyl geomteric methods \citep{Smolin:1979,Drechsler:1991,Drechsler/Tann,Drechlser:Higgs,Cheng:1988}. Most of these authors worked  at different places and separated from each other;  only in rare cases they knew and cited the respective works of their colleagues. In spite of the rising numbers of papers and authors  during the last third of the 20th century one cannot speak of the birth of a literary network or even of  a scientific subcommunity of Weyl geometrically oriented work in theoretical physics.    Once in a while also researchers studying gravitational gauge fields in the {\em Cartan geometric approach}  considered Weyl geometry as a special case of their wider program, with or without the additional feature of translational curvature; among them notably J. Charap, W. Tait, F.W. Hehl and E. Mielke \citep{Charap/Tait,Hehl_ea:1988Weyl_group,Hehl_ea:1988Kiel}. -- This list is, of course,  far from exhaustive.\footnote{For a  more extended, although still incomplete,  survey  see \citep{Scholz:2018Resurgence}.}

\subsection{Foundations of general relativity: EPS}
\label{subsec: EPS}
About the same time at which  Dirac and the Japanese physicists were using Weyl geometry in the context of scalar tensor theories new interest in Weyl's geometry of 1918 arose also in the research on the foundations of gravity. 
 Weyl's argument of \citep{Weyl:projektiv_konform} that the combination of projective and conformal structures  suffice for  uniquely   characterizing a Weyl geometric structure    was taken up, abstracted and extended  half a century later by {\em  J\"urgen Ehlers, Felix Pirani and Alfred Schild} (EPS in the sequel) \citep{EPS}. The EPS paper  was written for a {\em Festschrift} in the honour of J.L. Synge.
Synge was known for his proposal of basing general relativity  on the behaviour of standard clocks  rather than length measurements ({\em chronometric approach}).\footnote{Compare similar remarks in T. Sch\"ucker's contribution  to this volume.} From the foundational point of view, however, clocks are no less unproblematic for defining the metric of spacetime, because they are  realized by complicated material systems and would need a validated theory of time measurement for a foundational justification. The question suggested itself, whether the physical metric can be determined on the basis of more elementary  signal structures  of gravitational theory than material clocks, e.g. by like light rays, particle trajectories etc. Weyl's paper of 1921 had sketched  such a type of approach.

 EPS analysed Weyl's idea of 1921 in detail, using  the  mathematical language  of differentiable manifolds and mimicking,  at least to a certain degree,  Hilbert's axiomatic method. 
They started from three sets, $\mathcal{M}=\{p, q, \ldots\}, \, \mathcal{L}=\{L, N, \ldots\}, \,\mathcal{P}=\{P, Q, \ldots\} $    representing the collections of {\em events, light rays} and {\em particle trajectories} respectively. By postulates close to physical experimental concepts of light signal exchange between particles EPS formulated  different groups of axioms, which allowed them to conclude that the event set  $\mathcal{M}$ could be given the structure of a ($C^3$-) differentiable  manifold $M$ endowed with a  conformal structure $\mathfrak{c}$, the null-lines of which agree with $\mathcal{L}$, and a projective structure of ($C^2$-) differentiable paths in agreement with $\mathcal{P}$. The latter can equivalently be described  by a projective equivalence class $\mathfrak{p}= [\Pi]$ of affine connections $\Pi=\Pi^{\lambda}_{\mu \nu}$ (ibid. p. 77).
In an  additional axiom $C$ (ibid. p. 78) they secured the {\em compatibility} between the structures $\mathfrak{c}$  and  $\mathfrak{p}$, with the upshot ($C'$) that the null lines of the conformal cones of  $\mathfrak{c}$ are projective geodesics of $\mathfrak{p}$ (ibid. 78--80).\footnote{For a more detailed description of the paper, including references to follow up papers and some carefully critical remarks, see  \citep{Trautman:EPS}.} 

In the light of this result the criterion for  {\em compatibility of $\mathfrak{c}$ and $\mathfrak{p}$  in the sense of EPS} can be stated as follows:
\begin{quote}
A projective structure $\mathfrak{p}$ and a conformal structure  $\mathfrak{c}$   will be called {\em EPS-compatible}, if the null lines of $\mathfrak{c}$  are projective geodesics (autoparallels) of $\mathfrak{p}$. \label{Def EPS compatibility}
\end{quote}
This criterion was new. For Weyl the compatibility of $\mathfrak{p}$ and  $\mathfrak{c}$  was no independent problem, because he supposed both to be abstracted from a the affine connection of a Weylian metric (assumed to exist in advance). From the viewpoint of the EPS context a modernized definition of {\em Weyl compatibility}  can be stated as follows (compare sec. \ref{subsec: Inf Geo}):
 \begin{quote}
 {\em Definition}.  The conformal and projective structures ${\mathfrak{c}}$ and $ {\mathfrak{p}}$   are said to be {\em Weyl compatible}, if for some $g\in \mathfrak{c}$  a differential 1-form $\varphi$ can be found such that the  affine connection $\Gamma (g,\varphi)$ of the Weyl metric $[(g,\varphi)]$ satisfies $\Gamma (g,\varphi) \in \mathfrak{p}$.\footnote{If  this holds for {\em some} $g\in \mathfrak{c}$, then for {\em any} $g\in \mathfrak{c}$ (due to Weyl's rescaling result and the scale invariance of $\Gamma (g,\varphi)$).} \label{Def Weyl compatibility} 
 \end{quote}
In distinction to the Riemannian case $\varphi$  need, of course, not be integrable.

 Using their new concept of compatibility  EPS derived their main statement:
\begin{quote}
 A light ray structure $\mathcal{L}$ and a set of particle trajectories $\mathcal{P}$ defined on an event set 
$ \mathcal{M}$  which satisfy the EPS axioms  endow $ \mathcal{M}$ with the structure of a ($C^3$-) differentiable manifold $M$ and specify a  ($C^2$-) Weylian metric  $[ (g,\varphi ) ]$ upon $M$. The  metric is uniquely determined by the condition that its causal  and geodesic structures coincide with  $ \mathcal{L}$ and $ \mathcal{P}$ respectively.   
\end{quote}
 If one accepts the argumentation, this was an important result. 
The authors were  cautious enough, however, to  qualify their result by the remark:
 \begin{quote}
A fully rigorous formalization  has not yet been achieved, but we nevertheless hope that the
main line of reasoning will be intelligible and convincing to the sympathetic reader \citep[p. 69f.]{EPS}.
 \end{quote}
 In his commentary to a recent re-edition of the EPS paper Trautman hinted at the desideratum of a formal proof of the existence of the affine connection  $\Gamma (g,\varphi)$ (in our notation) \citep[p. 1584]{Trautman:EPS}. In a joint paper with Matveev the two authors even suggest to  give a counter example to EPS' existence claim \citep{Matveev/Trautman}; but this is due to the {\em quid pro quo} of Riemannian metrics and Weyl metrics (see sec.  \ref{subsec: Matveev/Trauman}). 
 
EPS gave arguments that assuming some physically  plausible axioms, spacetime may be described by a differentiable manifold $M$. Moreover, acccording to the authors  light rays and particle trajectories can be used to {\em establish} a metric on $M$ in the sense of {\em  Weyl geometry}, rather than assuming the latter  as a result of rod and clock measurements, chronometric  prescriptions, or even only as a structural property like Weyl had done in  \citep{Weyl:projektiv_konform}. 

Such a result was important also from the point of view of Einstein gravity; but  there remained a gap to the latter, namely how, or whether at all,  one can arrive at a (pseudo-) Riemanniang metric from the Weylian one. EPS argued that this gap might be closed
by a adding a  single additional {\em Riemannian axiom},  postulating the vanishing of  the scale curvature, $d\varphi=0$, i.e. the integrability of the Weylian metric \citep[p. 82]{EPS}. Such a postulate would not seem implausible, as Weyl's interpretation of the  scale connection $\varphi$ as electromagnetic potential, the {\em em}-dogma, was obsolete anyhow; of course EPS did not adhere to it. But the authors did not exclude the possibility that a scale connection field $\varphi$ of non-vanishing scale curvature might play the role of a ``true'', although still unknown,  field $f=d\varphi$ which would relate ``the gravitational field to another universally conserved current'' (\citep[p. 83]{EPS}).


The paper of Ehlers, Pirani and Schild triggered a  line of investigations in the foundations of general relativity, sometimes called the {\em causal inertial approach} (Coleman/Kort\'e), sometimes subsumed under the more general search for a {\em constructive axiomatics} of GRT (Majer/Schmidt, Audretsch, L\"a\-m\-merzahl, Perlick and others). The debate was opened by  \citep{Audretsch:1983}. It was soon continued by a collective  paper  written by  three authors  \citep{AGS}  and had  follow up studies, among others \citep{Audretsch/Laemmerzahl:1988}. The authors of the first mentioned  paper, Audretsch,  Straumann and G\"ahler,   argued that the ``gap'' between Weylian and Riemannian geometry  can ``be closed if quantum theory as a theory of matter is made part of the total scheme'' \citep[2872]{Audretsch:1983}. By this they referred to an investigation of scalar and spinor fields on a Weylian manifold and the assumption that the flow lines of their WKB-approximation coincide with the geodesics of the underlying metric. In this way these authors could underpin the  additional Riemannian postulate of EPS which was considered as sufficient for closing the gap between Weyl and Riemannian geometry.
 
But   the integrability of a Weylian metric is only a necessary condition for locally establishing  a Riemannian framework, not a sufficient one.  Although  the choice of the  Riemann gauge is always possible in an integrable Weyl geometry and appears natural from a purely mathematical point of view, the question remains which of the different scale gauges expresses  time  and other physical quantities directly. If it differs  from Riemann gauge, even the  conceptually minor  difference between the Riemannian structure and  an integrable Weylian structure  {\em matters for the physics of spacetime}.

The investigations of the causal inertial approach turned towards a  basic conceptual analysis from the point of view of foundations of inertial geometry \citep{Coleman/Korte:inertial_conformal}, some even looking for  Desargues type characterization of free fall lines \citep{Pfister:Newtons_law}.
 How  a kind of   ``standard clocks'' can be introduced in the Weyl geometric setting  without taking refuge to atomic processes, by   just  using the observation of light rays and inertial trajectories, was studied by 
\citep{Perlick:Diss,Perlick:1987,Perlick:Observerfields}.
Another line of  follow up works explored the extension of the foundational argument of the causal inertial approach to quantum physics, where  particle trajectories  might  no longer appear acceptable as a foundational concept.

\subsection{Geometrizing quantum mechanical configuration spaces }
\label{subsec: config space}
 A completely different perspective on Weyl geometry was taken by  {\em Enrico Santamato} in Naples.  
 In the 1980s  he proposed a new approach to quantum mechanics  based on studying   weak random processes of  ensembles of point particles  moving in a Weylian  modified configuration space \citep{Santamato:1984a,Santamato:1984b,Santamato:1985}. 
He compared his approach to that of   {\em Madelung-Bohm}  and to   the stochastic program of {\em Feyn\`es-Nelson}.\footnote{For E. Nelson's program  to re-derive the  quantum dynamics from classical stochastic processes and classical probability see \citep{Bacciagaluppi:Nelson}.}
While the latter dealt with stochastic (Brownian) processes,  Santamato's approach  was closer to the view of  Madelung and Bohm  because it assumed only random initial conditions, with classical trajectories  given in Hamilton-Jacobi form (this explains the attribute ``weak'' above). One can interpret  Bohm's particle trajectories as deviating  from those  expected in Newtonian mechanics by some ``quantum force''. Santamato found this an intriguing idea, but  he deplored its ``mysterious nature'' which   ``prevents carrying out a natural and acceptable theory along this line''. He  hoped to find a rational explanation  for the effects of the ``quantum force'' by means of a  geometry with a modified affine connection of the  system's configuration space. Then the deviation from classical mechanics would appear as the outcome of  ``fundamental properties of space''  \citep[p. 216]{Santamato:1984a},  understood in the sense of {\em configuration} space. 

In his first paper paper Santamato started from a configuration space with coordinates $(q^1, \ldots, q^n)$  endowed with a Euclidean metric. More generally,  his approach allowed for a general positive definite metric $g_{ij}$,  and later even a metric of indefinite signature,  for dealing with general coordinates of $n$-particle systems, and perhaps, in a further extension, with spin. The Lagrangian of the system, and the corresponding Hamilton-Jacobi equation, contained  the metric, either explicitly or implicitly.  This Euclidean, or more generally Riemannian, basic structure was complemented by a Weylian scale connection. Santamato's central idea was that the modification of the Hamilton-Jacobi equation induced by a properly determined {\em scale connection} could be used to  express the {\em  quantum modification}  of the classical Hamiltonian, much like was done in the  Madelung-Bohm approach. 
  Then the quantum aspects of the systems would be geometrized in terms of Weyl geometry, surely a striking and even beautiful idea, if it should work. 

Santamato thus headed towards a  new program of   {\em geometrical quantization sui generis}. This  had nothing to do with the better known geometric quantization program initiated more than a decade earlier by  J.-N. Souriau, B. Kostant and others, which was already well under way in the 1980s \citep{Souriau:1966,Kostant:1970,Simms:1978}. In the latter,  geometrical methods underlying  the canonical quantization were studied. Starting from a {\em symplectic  phase space} manifold of a classical system, the   observables were ``pre-quantized'' in a Hermitian line bundle, and  finally the Hilbert space representation of quantum mechanics was constructed on this basis.\footnote{See, e.g., \citep{Woodhouse:GeoQuant} or \citep[chaps. 22/23]{Hall:QuantumTheory}.
A classical monograph on the symplectic approach to {\em classical} mechanics is \citep{Abraham/Marsden};  but this does not discuss spinning particles.  In the 1980s the symplectic approach was already used as a starting platform for (pre-)quantization to which proper quantization procedures  could then hook up, see e.g. \citep{Sniatycki}.  Souriau was an early advocate of  this program. In his  book he   discussed relativistic particles with spin \citep[\S 14]{Souriau:1970}.} 
Santamato's geometrization was  built upon a different
 structure, Weyl geometry rather than symplectic geometry, and had rather different goals.

Like other proposals in the dBMB  (de Broglie-Madelung-Bohm) family, Santamato's program encountered little positive response.  In the following decades he 
  shifted  his research to  more empirically based studies in nonlinear optics of liquid crystals and quantum optics.
 Perhaps a critical paper by 
 {\em Carlos Castro Perelman}, a younger colleague who knew the program nearly from its beginnings, contributed to what would be an extended period of  interruption.  Castro discussed ``a series of technical points'' which seemed important for Santamato's program from the physical point of view \citep[p. 872]{Castro:1992}. This criticism may have contributed to a first interruption of work on the reearch program. It was taken up again and extended by Santamato, Martini and other researchers in the last decade.
 
 After the turn of the century, Santamato came back to foundational questions, working closely together with his  colleague {\em Franceso De Martini} from the University of Rome. Both had  cooperated in their work on quantum optics already  for many years. In the 2010s they turned  to geometrical quantization  in a series of joint publications that  continued the program Santamato had started  three decades earlier. They  showed how to deal with spinor fields in this framework, in particular with the Dirac equation \citep{DeMartini/Santamato:2013}, and they discussed the famous {\em Einstein-Podolsky-Rosen} (EPR) non-locality question \citep{DeMartini/Santamato:2014a,DeMartini/Santamato:2014b}. Moreover, they analyzed the helicity of elementary particles and showed that the {\em  spin-statistics relationship} of relativistic quantum mechanics can be derived in their framework without invoking arguments from quantum field theory \citep{DeMartini/Santamato:2015a,DeMartini/Santamato:2016a}.  In this new series of papers, the authors took  Minkowski space as the starting point for their construction  of the configuration spaces, which could be extended by internal degrees of freedom. They also enlarged the perspective by making a  transition from point dynamical Lagrangians to a dynamically equivalent  description in terms of  scale invariant  field  theoretic Lagrangians in two  scalar fields. 
For more details one may consult \citep{Scholz:2018Resurgence} from which this short survey  has been adapted. 

With these papers we have already entered  present work on Weyl geometric methods. In the next chapters we turn towards  actual researches in greater breadth.


\section{Interlude}
\label{sec: Interlude}
Let us first 
lay open the notations and concepts used in the rest of our  presentation. They are oriented at the historical literature but  deviate from it where it seems advisable. 

\subsection{Basic concepts, notation and IWG}
\label{subsec: Basic concepts}
 In the following we use scale weights such that quantities of dimension length  $L$ scale with weight 1,  $w(L)=1$ (different from Weyl's practice and, correspondingly,  the preceding part of this article in which the  convention of  ``quadratic'' scale weights with   $w(L)=\frac{1}{2}$ is used). Like usual we assume  a non-scaling vacuum velocity of light $c$ and a non-scaling  Planck constant $\hbar$;  time $T$, mass $M$, and energy $E$  scale then respectively by  $w(T)=1, \; w(M)=w(E)=-1$.\footnote{If F. Hehl's suspicion that the velocity of light  may vary with the gravitational potential turns out to be right (if one considers more precise solutions of the Maxwell equation in GR than assumed in the  optical limit), a specification like, e.g., a constraint to conformally flat Riemannian metric, has  to be included  in  the definition of $c$.  } 
 These weights will be called {\em Weyl weights}; they are the negatives of the mass/energy weights used by elementary particle physicists.
 
 The simplest way to define a {\em Weylian metric} on a differentiable manifold $M$ is by specifying an equivalence class $\mathfrak{g}=[(g,\varphi)]$, consisting of a pseudo-Riemannian metric $g$ (locally $g=g_{\mu \nu}dx^{\mu}dx^{\nu}$), the {\em Riemannian component} of the Weylian metric,  and a real-valued differential 1-form $\varphi$ (locally $\varphi= \varphi_{\nu}dx^{\nu}$) representing the {\em scale connection}. Equivalence $(g,\varphi) \sim (\tilde{g},\widetilde{\varphi})$ is defined by the gauge transformation
 \beq \tilde{g}= \Omega^2 g, \qquad 
 \widetilde{\varphi} = \varphi - d \log \Omega = \varphi - \frac{d \Omega}{\Omega}\, , \label{gauge trafo}
 \eeq
with a strictly positive re-scaling function $\Omega$ on $M$.\footnote{Note the difference to eq. (\ref{eq gauge trafo}), due to the different conventions for scaling weights.} Choosing a representative $(g,\varphi) \in \mathfrak{g}$ means to {\em gauge} the metric. 

If the scale connection is closed in some gauge $(g,\varphi)$ , $d \varphi=0$, it is so in any gauge and is locally (i.e., in simply connected regions) exact with potential, say $-\omega$, such that $\varphi= - d \omega$. Weyl's length transfer can then be  integrated 
\[ \Omega = e^{\int_{\gamma}\varphi(\dot{\gamma})}=e^{-\omega}
\] 
independently of the path $\gamma$. One therefore speaks of an {\em integrable Weyl geometry} (IWG). Locally a  gauge of the form $(\tilde{g},0)$ with
\beq \tilde{g}=  e^{2\int_{\gamma}\varphi(\dot{\gamma})}= e^{-2\omega}g \quad \mbox{and} \quad \tilde{\varphi}= \varphi + d\omega =0 \label{Riemann gauge}
\eeq 
can be chosen, the {\em Riemann gauge} of an IWG.   Because of the remaining freedom of choosing different scale gauges {\em this does not mean}, however,  a  structural {\em reduction to Riemannian geometry}. That may be  important for physical applications, if there are reasons to assume or to explore the possibility that measuring instruments (``clocks'') don't adapt to the Riemann gauge. The Jordan frame of  Brans-Dicke theory, e.g., corresponds to Riemann gauge if the latter is being formulated in terms of IWG. The still ongoing debate on the question which frame, respectively  gauge, expresses measured quantities best in BD theory shows that an a priori  preference for the Riemann gauge  may be a shortcut shadowing a physically important  question. 

A more fundamental difference to Riemannian geometry arises, if there is an obstruction $f= d\varphi \neq 0$. $f$ is the  curvature of the scale connection, shorter the {\em scale curvature} (not to be confused with the scalar curvature $R$ of the affine connection). Physically it expresses the field strength of the scale connection, often called the {\em Weyl field} of the structure.

 The uniquely defined {\em affine connection} compatible with  $[(g,\varphi)]$ will often be written as $\Gamma$, if context makes clear what is meant. In more complicated contexts we write more explicitly $\Gamma=\Gamma\,(g,\varphi)$, although the second expression might suggest a scale gauge dependence of $\Gamma$, which is not the case.   $\Gamma$ decomposes into gauge dependent contributions like in (\ref{eq affine connection}), $\Gamma\,(g,\varphi)= \Gamma = \Gamma\,(g)+\Gamma\,(\varphi)$ with $\Gamma(g)$ the Levi-Civita connection of $g$ and 
 \beq \Gamma^{\mu}_{\nu \lambda}(\varphi)=  \delta^{\mu}_{\nu} \varphi_{\lambda} +  \delta^{\mu}_{\lambda} \varphi_{\nu} - g_{\nu \lambda}\,\varphi^{\mu} \, , \label{Gamma(varphi)}
 \eeq
  here without the factor $\frac{1}{2}$ in comparison to (\ref{eq affine connection})  because of the modified weight convention.  $\Gamma$ itself is gauge invariant.
 
 The covariant derivative $\nabla$ and the curvature tensors $\mathit{Riem}, \, \mathit{Ric}$ will usually  denote the ones due to the Weyl geometric affine connection $\Gamma$. They are scale invariant. In any gauge they decompose into a gauge dependent Riemannian component,   $\mathit{Riem}\, (g)$, $\mathit{Ric}\, (g)$, and another one expressing the contribution of the scale connection  $\mathit{Riem}\, (\varphi), \mathit{Ric}\, (\varphi)$, etc.\footnote{Compare fn. \ref{fn curvature conventions}. For the scale connection component $\mathit{Riem}\,(\varphi)$  of the Weyl geometric Riemann curvature one finds  $R(\varphi)^{\mu}_{\;\; \nu \lambda \kappa}=  \nabla(g)_{\lambda}\, \Gamma^{\mu}_{\nu \kappa}  - \nabla(g)_{\kappa} \, \Gamma^{\mu}_{\nu \lambda}  +  \Gamma^{\alpha}_{\nu \kappa} \Gamma^{\mu}_{\alpha \lambda}  - \Gamma^{\alpha}_{\nu \lambda} \Gamma^{\mu}_{\alpha \kappa}  $ \citep[eq. (9)]{Yuan/Huang:2013}.  
 There is also a gauge invariant decomposition of $\mathit{Riem}$ into what Weyl called directional curvature and length curvature, which must not be confused with the gauge dependent decomposition just mentioned. The {\em length  curvature} $F$ as part of the Riemann tensor is closely related to the scale connection 2-form $f=d\varphi$; its   components are $F^{\mu}_{\nu \kappa \lambda}=- \delta^{\mu}_{\nu}f_{\kappa \lambda}$. As $F$ and $f$ can easily be translated into one another;  they are often identified in loose speech.
 }
 The scalar curvature scales with weight $w(R)=-2$. 

The scaling behavior of  (scalar, vector, tensor, spinor) fields $X$  on a Weylian manifold $(M, [(g,\varphi)]$ has to be specified by a corresponding scale weight $w(X)$ according to invariance principles of the Lagrangian or, from a more empirical point of view, according to its physical dimension. From a mathematical point of view $w(X)$ specifies the representation type of the scale group for the adjoint bundle (to the scale bundle) in which $X$ lives.

The full covariant derivative of fields $X$ in the sense of the Weyl geometric structure in a gauge $(g,\varphi)$ has to take  into account the scale connection in addition to the Levi-Civita of $g$. It will be called {\em scale covariant derivative} $D$ (Dirac's ``co-covariant derivative, p. \pageref{Dirac co-co}) and is given by
\beq DX = \nabla(g)\,X +w(X) \otimes X \, . \label{scale covariant D}
\eeq 
In coordinates this is for a vector field $\xi$ of weight $w$, e.g.,   $D_{\mu} \xi^{\nu}= (\nabla_{\mu}+ w \varphi_{\mu})\, \xi^{\nu}$. Weyl understood the compatibility of a Weylian metric represented by $(g,\varphi)$ with an affine connection $\Gamma$ in the sense that parallel transport by $\Gamma$ has to ``respect the length transfer''. This condition can be stated more formally as vanishing of the scale covariant derivative of $g$:
\beq D g = 0 \quad \Longleftrightarrow \quad \nabla(\Gamma)\, g = - 2 \varphi \otimes g \label{compatibility metric}
\eeq
The term $Q$ with $Q_{\lambda \mu \nu } := -2 \varphi_{\lambda} g_{\mu \nu}$ on the r.h.s. of the equation is a  necessary conceptual feature for ensuring the {\em  metric compatibility of the affine connection} in the sense of Weyl geometry. 
From the viewpoint of Riemannian geometry, in contrast, it  appears as a deviation from the norm, and has been termed ``non-metricity'' by Schouten,  in this special form containing $g$ as ``semi-metricity''. Using the scale covariant derivative the Weyl geometric affine connection can be written in a form close to the one for the Levi-Civita connection  \citep[eq. (46)]{Jimenez/Koivisto:2014}:
\beq 
\Gamma(g,\varphi)^{\lambda}_{\mu \nu} = \frac{1}{2}g^{\lambda \alpha}(D_{\mu}g_{\nu \alpha}  +D_{\nu}g_{\alpha \mu}  -  D_{\alpha}g_{\mu \nu}  )
\eeq

The condition (\ref{compatibility metric}) is sometimes used as an entry point for defining Weyl structures (pseudo-Riemannian, Hermitian, quaternionic etc.).
\begin{quote}
Definition: A (pseudo-Riemannian) {\em Weyl structure} is given by a triple $(M, \mathfrak{c}, \nabla)$ of a differentiable manifold $M$, a pseudo-Riemannian conformal structure $\mathfrak{c}=[g]$ and a covariant derivative $\nabla$, such that  for all ${g \in \mathfrak{c}}$ there is a differential 1-form ${\varphi}$  for which $ \nabla g + 2 \varphi\otimes g = 0$.
\end{quote}
Gauge transformations of type (\ref{gauge trafo}) for changes of the representative of $\mathfrak{c}$ are a consequence of this definition \citep{Ornea:2001}. 

Weyl introduced {\em scale invariant geodesics} by the usual condition $\nabla_{u}\, u=0$  for the covariant derivative $\nabla=\nabla(g,\varphi)$ of the tangent field $u=\dot{\gamma}(\tau)$ of a curve $\gamma(\tau)$. In coordinates and with $\Gamma=\Gamma(g,\varphi)$ this is
\beq
\ddot{\gamma}^{\lambda} +\Gamma_{\mu\nu}^{\lambda} \dot{\gamma}^{\mu}\dot{\gamma}^{\nu}=0  \, .\label{scale invariant geodesic}
\eeq
Different to Riemannian geometry, scale invariant geodesics are (usually) not parametrized by curve length. This can be changed by introducing  gauge dependent re-parametrizations of $\gamma$, that is a class of curves with the same trace as Weyl's scale invariant geodesics such that the tangent fields scale with weight $w(\dot{\gamma})=-1$. This leads to introducing 
 {\em scale covariant geodesics} of weight $w(\dot{\gamma}) =-1$, characterised  by the condition  $D_{\dot{\gamma}}\dot{\gamma}=  \nabla_{\dot{\gamma}}\, \dot{\gamma} - \varphi( \dot{\gamma})\, \dot{\gamma}=0$, or in coordinates
 \beq
 \ddot{\gamma}^{\lambda} +\Gamma_{\mu\nu}^{\lambda} \dot{\gamma}^{\mu}\dot{\gamma}^{\nu} -\varphi_{\nu}\dot{\gamma}^{\nu}  \dot{\gamma}^{\lambda}   =0    \, .\label{scale covariant geodesic}
 \eeq 
The  difference  between the solutions of  (\ref{scale invariant geodesic}) and (\ref{scale covariant geodesic}) (in coordinates  $\varphi_{\nu}\dot{\gamma}^{\nu} \dot{\gamma}^{\lambda} $ ) is proportional to $\dot{\gamma}$. It thus  does not change the direction of the curve, but only  its parametrization. This results in $\nabla_{\dot{\gamma}}(g(\dot{\gamma},\dot{\gamma}))=0 $ for a scale covariant geodesic,  and a  constant length function $l= g(\dot{\gamma},\dot{\gamma})$ like in Riemannian geometry.\footnote{For the Riemannian case see, e.g., \citep[p. 92]{Kuehnel}.} 

The length function $l$  is not only constant along a scale covariant geodesic but also scale invariant, because of $w(l)=w(g(\dot{\gamma}, \dot{\gamma}))= 2-1-1=0$. For $l=1$ the  integrated length of a geodesic (with regard to the Riemannian $g$ component of a gauge $(g,\varphi)$) can be read off from the parametrization. It is, of course, scale dependent.

Similarly the (squared) {\em Riemannian length} of any differentiable curve $ l_R^2(\gamma)= \int g(\dot{\gamma(\tau)},\dot{\gamma(\tau)})d\tau$ is a scale dependent quantity. If one corrects the integrand  by the Weylian length transfer function, one finds that the integral
\beq l_W^2(\gamma) = \int e^{2 \int \varphi(\dot{\gamma})} g(\dot{\gamma(\tau)},\dot{\gamma(\tau)})d\tau
\eeq 
leads to a scale invariant curve length $l_W$ (up  to a constant factor depending on the value of the rescaling function $\Omega$ at the initial point of the curve  $e^{\Omega(\tau_0)}$).  This holds   for any $\varphi$. It can be considered as the {\em Weylian length} of the curve.   For an integrable Weyl geometry this boils down to measuring curve lengths in Riemann gauge  (\ref{Riemann gauge}).

\subsection{Weyl geometric gravity, in particular in IWG }
\label{subsec: Weyl geometric gravity}
The following remarks refer to the gravitational sector  only. Theories building on such an approach will add matter fields and their couplings to the gravitational sector (some special case follow in the next sections). Scalar fields  coupling to the Hilbert term will be treated here as part of the gravitational sector, although in certain respects, in particular their energy-momentum, they also behave like matter fields. A matter-like contribution to the energy-momentum results  from their kinetic term   which is not considered in this section, but  see sec. \ref{subsec: cosmology and dm}.

 Weyl geometric gravity is a scale in/co-variant theory. Its 
Lagrangians $L$  are {\em scale covariant} functions of weight $w(L)=-n$ 
with a {\em scale invariant} Lagrange density $\mathfrak{L}=L\sqrt{|det\,g|}$.
 In the following we assume $n=4$. Typical gravitational Lagrangians are  then of the form
\beq  {L}_{\mathit{grav}} =  \alpha_1 \phi^2 R + \alpha_2 f_{\mu \nu}f^{\mu \nu}+ \alpha_3 \phi^4 +\beta _1  R^2 +\beta _2\, \mathit{Ric}^2 +\beta_2\, \mathit{Riem}^2 + \ldots \, ,\label{L-grav general}
\eeq
where $\phi$ is a scale covariant real scalar field of weight $w(\phi)=-1$, $\mathit{Riem}, \; \mathit{Ric}$, and $R$ denote the Weyl geometric Riemann, Ricci and scalar curvatures, $f= d\varphi$ is the field strength of the scale connection (Weyl field), $\mathit{Riem}^2=R^{\mu}_{\;\;\nu \kappa \lambda}R_{\mu}^{\;\;\nu \kappa \lambda}, \; \mathit{Ric}^2=R_{\mu \nu}R^{\mu \nu}$,  and the dots indicate other quadratic terms in Weyl geometric cuvature of weight $-4$ including, if one wants, the (conformal) Weyl tensor of the Riemannian component in any gauge. Weyl never considered the $\alpha_1$ and $\beta_2$ terms (i.e., for him always  $\alpha_1=\beta_2=0$), moreover he considered $\beta_1=0$ in \citep{Weyl:GuE}, respectively  $\beta_3=0$  in \citep{Weyl:RZM5}; Dirac assumed all $\beta_i=0\; (i=1,2,3)$  etc.
There are indications that the quadratic curvature terms of the classical Lagrangian and the Weyl field term  may be of importance in strong gravity regimes and  also for the quantization of gravity (sec. \ref{subsec: SM and grav}, \ref{subsec: cosmology and dm}).

 With $\beta_1=\beta_3= 1, \beta_2=-4$ the explicitly given quadratic curvature terms  of  (\ref{L-grav general}) in $\mathfrak{L}$ reduce to the scalar density $\kappa$ of the Gauss-Bonnet theorem as  generalized by  Chern: 
 \beq \kappa= (R^2 -4  \mathit{Ric}^2 + \mathit{Riem}^2) \sqrt{|det\,g|}  \label{Gauss-Bonnet constraint}
 \eeq 
 For orientable compact differentiable manifolds 
 in dimension $n=4$ the theorem tells us that
\[ 32 \pi \,  \chi(M)= \int_M \kappa(g)\, dx \, ,
\]
with $\chi(M)$ the Euler-characteristic of $M$ and $g$ any Riemannian metric $g$ on $M$. Interestingly recent authors realized that Weyl geometric field theories for which the Riemannian component of the quadratic curvature Lagrange densities reduce to the  topologically preferred  Gauss-Bonnet form $\kappa dx$ behave particularly well with regard to  unitarity \citep{Tanhayi_ea:2012} and stability \citep{Jimenez/Koivisto:2014}.

Scale covariant theories with $\alpha_1\neq 0$ and a nowhere vanishing gravitational scalar field  have a scale gauge in which the scalar field is constant, $\phi \doteq \phi_0$. Here the following  convention is used: 
\begin{quote}
$\doteq \quad$ denotes equalities which only hold is specific scale gauges.
\end{quote}
The gauge with $\phi \doteq \phi_0$ will be called the {\em scalar field gauge} with regard to $\phi$ (in short $\phi$-gauge). If the coefficient of the modified Hilbert term  assumes the value, $\alpha_1 \phi^2 \doteq (16\pi G)^{-1} \sim \frac{1}{2} E_p^2$ ($E_P$ the reduced Planck energy), it specifies  the {\em Einstein gauge}.
Theories with  a distinguished scalar field $\phi$ allow to define:
\begin{quote}
 With regard to $\phi$    {\em  scale invariant observable quantities}  
   of a scale covariant field $X$ with weight   $w=w(X)$ are given by forming the proportion with $\phi^{-w}$, i.e, by  $ \check{X} = \phi^{w}X $.
\end{quote}

This boils down to determining the field values in $\phi$-gauge (up to the constant factor $\phi_o^w$). Probably that was the reason for Utiyama to consider $\phi$ as a ``measuring field'' (see p. \pageref{Utiyama}).
These theoretical considerations may get underpinned physically, if    reasons turn up which allow to conclude that the result of measuring processes (e.g., atomic clocks) are displayed in the  scalar field gauge of $\phi$. The scaling of the Higgs field $\Phi$  indicates in this direction, if it underlies a common biquadratic potential with the gravitational field $\phi$ (sec. \ref{subsec: SM and grav}).

If one spells out the modified Hilbert term of (\ref{L-grav general}) in terms of the Riemannian and the scale connection contributions to $R$,
\beq R= R(g)+R(\varphi) \quad \mbox{with} \quad R(\varphi)=-(n-1)(n-2)\varphi_{\nu}\varphi^{\nu} - 2(n-1) \nabla(g)_{\nu}\, \varphi^{\nu}
\eeq 
 a mass term $\frac{1}{2}m_{\varphi}^2 \varphi_{\nu}\varphi^{\nu}$ for the Weyl field turns up, which in dimension $n=4$ would indicate the impressive value  $m_{\varphi}^2= 6 \,E_P^2$. Even if one allows for quantum corrections, or includes other terms in an approach of this type, one arrives at a mass far beyond experimental access (e.g. at the  LHC).
Bosonic couplings of the  Weyl field thus occur at extremely small distances (close to the Planck length); their effects integrate out at longer distances, comparable to the invisibility of the weak interaction at microscopic scales.

 For investigations in {\em low energy regions}, i.e., at laboratory or astronomical scales, Weyl geometric gravity can thus be  {\em effectively }  {\em modelled } in terms of scale connections with vanishing curvature, $d  \varphi=0$, i.e. in  the  framework of {\em integrable Weyl geometry}. In a Lagrangian approach the reduction to the effective theory can be expressed  by introducing Lagrange multiplier terms $\lambda_{\mu \nu}f^{\mu \nu}$ 
(like in (\ref{Lagrange constraint f=0})). In the following we therefore introduce a simplified version of Weyl geometric gravity in the  framework of IWG by  the external constraint of a vanishing scale curvature, $d\varphi=0$. Even then the  scalar field $\phi$ contains an  additional degree of freedom in comparison with Einstein gravity. In the Riemann gauge it can be expressed as  $\phi \underset{Rg}\doteq \phi_0 e^{-\sigma}$, 
 where $\underset{Rg}\doteq$  denotes equality in the Riemann  gauge; similarly   $\underset{Eg}\doteq$ for the Einstein gauge. 
 
 The simplified version  of Weyl geometric gravity is   a  {\em scalar-tensor theory} in which the new scalar degree of freedom is expressed partially by the scalar field and partially by the (integrable) scale connection. This can be seen as follows:  
 The data in an arbitrary scale gauge arise from Riemann gauge by  (length-) rescaling with a real valued function $\Omega=e^{\omega}$:
\beq g \doteq e^{2\omega}g , \qquad 
 \quad \varphi_{\nu} \doteq-\partial_{\nu} \omega \quad (\varphi \doteq - d \omega ),  \qquad
\phi \doteq \phi_0\, e^{-(\sigma + \omega)}
\eeq
The information of the new dynamical degree of freedom is now no longer  encoded by the scalar field alone, but is being distributed among the scalar field and the scale connection, i.e., encoded by the pair $(\phi,\varphi)$.
In the Einstein gauge $(\hat{g},\hat{\varphi}, \hat{\phi})$, specified by  $\hat{\phi}\doteq\phi_0$, the respective values are
\beq \hat{g} \underset{Eg}\doteq e^{-2 \sigma}g \, \qquad 
\hat{\varphi}_{\nu} \underset{Eg}\doteq \partial_{\nu}\sigma\, \qquad 
\hat{\phi} \underset{Eg}\doteq \phi_0 \quad \mbox{(constant)} \, . \label{eq Einstein gauge}
\eeq 
The dynamical information of the new degree of freedom is here completely contained in the integrable scale connection with integral $\sigma$, while the scalar field has become a constant.   This has interesting dynamical consequences for inertial paths, if they follow Weyl geometric geodesics  (section \ref{subsec: cosmology and dm}). 

 

At least test bodies do so. 
This is the case, because the doubly covariant energy momentum tensor $T_{\mu \nu}$  scales with weight $-2$ (also in the non-integrable case).\footnote{For the Hilbert energy tensor $ T^{(m)}_{\mu \nu}= - \frac{2}{\sqrt{|g|}}\frac{\delta \mathfrak{L}_{m}}{\delta g^{\mu\nu}}=-\left(2 \frac{\partial L_m}{\partial g^{\mu\nu}}-L_m\, g_{\mu\nu}\right) $ this results  from the scaling weights $w(L_m)=-4$ and of $g$ (similarly for the canonical energy tensor). It  agrees with the ``phenomenological'' weight read off from  the physical dimension of the energy density $T^0_{\;\;0}$; in dimension 4 $ w(T^0_0)=w([EL^{-3}])=-4$.  }
The ``conservation'' condition  of GR 
 translates, at least in IWG,  to  a vanishing scale covariant divergence
\beq D_{\mu}T^{\mu}_{\;\; \nu} = 0 \, .
\eeq
That allows to generalize 
 the classical Geroch-Jang theorem establishing the {\em geodesic principle} in  Einstein gravity {\em for test bodies}  \citep{Geroch/Jang:1975}  without great effort  to the {\em framework of integrable Weyl geometry}  \citep[app. 5.3]{Scholz:2019WST}. 
 The case of extended bodies is more complicated and needs further clarification.\footnote{The detailed study in \citep{Ohanian:2010} discusses, among others, the problem for the geodesic principle of extended bodies in BD theory. This cannot be translated $1:1$ to the IWG case. }

Let us be content with these short remarks. In the next section  some  topics in the foundations of gravity  and Weyl geometric researches in  other subfields  of physics will be  reviewed. 

\section{Interest today}
\label{sec: Today}
After  the reinterpretation of the original gauge idea as a phase factor in quantum mechanics in the second half of the 1920s Weyl's  scale geometry (purely infinitesimal geometry) of 1918 lay dormant for nearly half a century, enhanced by Weyl's own clear disassociation from this idea in the 1940s. Surprisingly, it lived up again rejuvenated in the early 1970s,  enriched by new concepts and triggered by new interests in conformal transformations in gravity theory, cosmology, quantum physics, and field theory. The following section   discusses selected  topics in which Weyl geometric methods enter crucially into ongoing present research. Of course this selection is quite subjective; it  is necessarily  biased by my own  interests and delimited by the restrictions  in perspective and knowledge of the author.

\subsection{Philosophical reflections on gravity}
\label{subsec: Philosophical reflections}
Extending the Riemannian framework for gravity theories may help to clarify conceptual questions. This is, of course, a truism and holds not only for the Weyl geometric generalization but also, e.g., for Cartan geometry. Her we deal with the first option. A classical subject for this type of questions is the discussion of the problem which is the ``physical frame''  in Brans-Dicke theory.

 Brans-Dicke theory presupposes the framework of  Riemannian geometry and uses conformal transformations not only for the metric but also for the scalar field, while keeping the affine connection arising as a solution of eq. (\ref{L_JBD}) fixed.\footnote{For modern presentations of BD theory see, e.g. \citep{Fujii/Maeda}; the transformation  of the Levi-Civita geodesics of the Jordan frame to other frames are discussed in (ibid, sec. 3.4).} 
In a technical sense it  works in the framework of an integrable Weyl structure $(M, \mathfrak{c}, \nabla)$, where the Jordan frame corresponds to the Riemann gauge of the Weylian metric.  For most practitioners this goes unnoticed. Exceptions are \citep{Quiros_ea:2013} and the authors of the Brazilian group of Weyl geometric gravity, \citep{Almeida/Pucheu:2014,Almeida/Pucheu/Romero:2014,Fonseca/Romero_ea:GR_Weyl_frames}.\footnote{Note, however, that Romero et al. have a peculiar  view of BD theory, different from the usual one and influenced by their  paradigm  called WIST ({\em Weyl integrable spacetime}). This is  a theory of gravity placed in an integrable Weyl geometric framework, but with a broken scale symmetry and Levi-Civita geodesics of the distinguished frame. In their view also in  BD the free fall trajectories are determined  by the Levi-Civita geodesics of a distinguished frame (which can be the Einstein frame)  \citep[sec. 6]{Romero_ea:2012GR}. This modified version of BD theory is  equivalent to WIST \citep[sec. 8]{Romero_ea:2012GR}. \label{fn WIST}}
This theory is  nearly  
a Weyl geometric scalar tensor theory arising from (\ref{L-grav general}) with $\alpha_2=\beta_i=0, \; i=1,2,3$ by adding a quadratic scale covariant kinetic term of the scalar field
\beq L_{\partial \phi2} = -\frac{\alpha}{2} D_{\nu}\phi\, D^{\nu}\phi \, ,
\eeq
even though the $L_{\partial \phi2}$  is usually not written explicitly in a scale covariant form but given in the Jordan frame in the form  $ L_{\partial \phi2} \doteq  -\frac{\alpha}{2} \partial_{\nu}\,\phi  \partial^{\nu}\phi$. Additional terms arise in other frames from the respective conformal transformations. I consider such an approach as {\em nearly} Weyl geometric gravity, because often a matter term $L_m$ is added, which  breaks the scale invariance of the Lagrange density. That raises the question in which scale gauge (frame)  the matter Lagrangian couples minimally to the Riemannian component of the metric. This is usually called the {\em physical metric} (or frame) \citep[p. 227f.]{Fujii/Maeda} and one has to decide in which relation the physical metric/frame/gauge  stands to the Riemann/Jordan frame and the Einstein frame. In a Weyl geometric scalar tensor theory as conceived here, one better  writes  the matter Lagrangian in a scale covariant form ($w(L_m)=-4$). This leads to scale  {\em invariant dynamical equations} (containing scale {\em co}variant terms). 

 The physical breaking of the scale symmetry is a separate question and can be discussed at a later stage of the theory development. It is related to fixing the   basic parameters of the measuring  processes.\footnote{Cf.  \citep{Quiros_ea:2013} and  \citep{Scholz:Paving}.}
Then one arrives at the scale gauge in which observational values are expressed directly. Let us assume, for the moment, that this is, e.g.,  the Einstein gauge.\footnote{For an argument in favour of the Einstein gauge (under certain assumptions) see section \ref{subsec: SM and grav},  observation ($\ast$). } 
This is then also the frame (or gauge) in which the Lagrangian matter terms are expressed most directly in the form known from non-scale invariant formulations. 
Insofar this is the ``physical gauge''. The Riemann gauge has, on the other hand, the pleasant feature that the Weyl geodesic geodesics are identical to the Levi-Civita ones of the gauge. From a more conservative  viewpoint one might  say that the free fall trajectories of BD gravity are those of Riemann gauge. But free fall and matter Lagrangian are both important physical features of a  theory of gravity. Which one should we then consider as ``physical''? The {\em two frames  express different aspects of physical reality} in a mathematically most direct way. In this sense both can be considered as ``physical''; it does not matter which way, we only need to be clear what is meant.

C. Romero, M. Pucheu and other authors of the Brazilian group mentioned above (sec. \ref{subsec: WST}) use the framework of  Weyl geometric gravity  to discuss different  inter-theory relations \citep{Romero_ea:2012GR,Romero_ea:2012flat}. In the first paper they discuss how Einstein gravity can be expressed in this  wider framework by a  scale-covariantization of dimensional quantities.  In particular the gravitational constant is promoted to a scalar field (proportional to $\phi^2$ in our notation), which also couples to the matter term (which then becomes $ \phi^4 L_m$ ). Their version of Weyl geometric scalar tensor theory (WIST) is shown to be equivalent to their view of BD theory.\footnote{See fn. \ref{fn WIST}.} 
 Finally they discuss the relationship between original  Nordstr\"om's scalar theory of gravity  to the Einstein-Fokker version from the point of view of conformal transformations of a flat Lorentzian manifolds  \citep{Romero_ea:2012flat}. This has been taken up in a recent broader philosophical discussion of the usefulness of a Weyl geometric point of view for this type of theory relation by P. \citet{Duerr:2019}.

\subsection{Matveev/Trautman on EPS}
\label{subsec: Matveev/Trauman}

A critical point of the EPS approach, the question whether EPS compatibilty of a  projective and a conformal structure implies Weyl compatibility, has recently been scrutinized by \citet{Matveev/Trautman}. 
 Weyl presupposed the compatibility of a given pair of conformal and projective structures by considering only  pairs which arise from a (presupposed) Weylian metric. 
  In the EPS approach the compatibility question  has to be posed in greater generality;  it was formulated in terms of the  geometrical criterion of EPS-compatibility  stated above (p. \pageref{Def EPS compatibility}).  Clearly Weyl compatibility implies EPS-compatibility. The main result of EPS claims the inverse.
  
   A counter-example to the equivalence of the two compatibility  criteria would falsify the proof argumentation  for the main result of EPS.  Referring to Weyl and the EPS paper,  Matveev and Trautman  announce   ``a theorem giving the necessary and sufficient conditions for compatibility of conformal and projective structures''  and proudly claim that their result \citep[p. 822]{Matveev/Trautman} ``completes a line of research initiated by Weyl and continued by physicists'' (ibid. p. 824). They do so by giving a simple example of a non-integrable Weyl structure  which they consider as a counterexample to the EPS result (ibid., sec. 3.). Of course this is wrong. 
 
 The erroneous claim of having invalidated the main result of EPS results from a simple {\em quid pro quo}. 
  The authors of  \citep{Matveev/Trautman} use a compatibility criterion of the structures ${\mathfrak{p}}$ and $ {\mathfrak{c}}$ relying on the framework of Riemannian geometry and thus   
  different from Weyl and from EPS:
   \begin{quote}
   {\em Definition.} The conformal and projective structures ${\mathfrak{c}}$ and $ {\mathfrak{p}}$   are said to be compatible if there is $ g \in \mathfrak{c}$ such that $\Gamma(g) \in  {\mathfrak{p}}$. \citep[p. 823, notation slightly adapted]{Matveev/Trautman}
\end{quote}    
 This definition   restricts the underlying metrical structure to  the Riemannian case.  I therefore call   it the {\em Riemann compatibility} of a conformal and a projective structure. 
  

 Of course Riemann compatibility  is a much stronger criterion than Weyl compatibility (p. \pageref{Def Weyl compatibility});  no great wonder that  Matveev/Trautman can give ``counter examples'' of pairs $({\mathfrak{p}},{\mathfrak{c}})$  for which the EPS compatibility holds, but not  their own one.

A veritable achievement of the paper lies, on the other hand, in formulating  an analytical criterion  for the compatibility of  two structures $\mathfrak{p}$ and $\mathfrak{c}$  (p. 823, ibid.), 
even though it is established for  Riemann 
 compatibility  only \citep[equs. (7), (8)]{Matveev/Trautman}. It can easily  be  generalized (i.e. weakened)  to the case of Weyl compatibility in the following way.
 
For given ${\mathfrak{p}}$ and ${\mathfrak{c}}$   
first choose an affine connection  $\overline{\Gamma}\in \mathfrak{p}$ and a Riemannian metric 
$g\in \mathfrak{c}$, the latter with  Levi-Civita connection $\Gamma(g)$. Any Weylian metric  subordinate to the conformal structure  $\mathfrak{c}$ can be expressed  in a scale gauge with Riemannian component $g$, i.e. in a  gauge of the form  $(g, \varphi)$ with some real differential form $\varphi$. Denote its (Weylian) affine connection by $\Gamma(g,\varphi)$.

Using the Thomas symbol $\Pi(\Gamma)$ of an affine connection $\Gamma$ 
 \[ \Pi^{i}_{jk}(\Gamma)= \Gamma^i_{jk}-\frac{1}{n+1}\delta^i_j\Gamma^l_{lk}-\frac{1}{n+1}\delta^i_k\Gamma^l_{lj} \, 
\] 
a necessary and sufficient condition for the projective equivalence of two affine connections $\Gamma$ and $\Gamma'$ is the equality of their Thomas symbols, respectively  the vanishing of the Thomas symbol for the difference:  
\[ \Pi(\Gamma) - \Pi(\Gamma') = \Pi(\Gamma -\Gamma')=0 \]
For checking the Weyl compatibility of ${\mathfrak{p}}$ and ${\mathfrak{c}}$  one has to see, whether $\Gamma(g,\varphi)$ and  $\overline{\Gamma}$ are projectively equivalent, i.e. whether or not the Thomas symbol of their difference vanishes.

Calculating first the Thomas symbol  $T$ for the difference of the Levi-Civita connection of $g$ and of  $ \overline{\Gamma}$, 
\beq   T(g,\overline{\Gamma}) := \Pi (\Gamma(g) - \overline{\Gamma})\, ,
 \eeq
the searched for difference  $\Pi (\Gamma(g,\varphi) - \overline{\Gamma}) $   can be expressed by $T(g,\overline{\Gamma})$ and $\varphi$; it will be denoted by  $\widetilde{T}(g,\varphi,\overline{\Gamma}) = \Pi (\Gamma(g,\varphi) - \overline{\Gamma})$. In  components it is:
 \beq \widetilde{T}(g, \varphi, \overline{\Gamma})^i_{jk}=T(g,\overline{\Gamma})^i_{jk}+\frac{1}{n+1}( \delta^i_j\varphi_k+\delta^i_k \varphi_j)-g_{jk}\varphi^i \,  \label{eq Matveev/Tratuman tensor}
 \eeq 
Vanishing of $\tilde{T}$ is necessary and sufficient for the projective equivalence of $\Gamma(g,\varphi)$ and $\overline{\Gamma} $, and thus the Weyl compatibility of the given projective and conformal structures.\footnote{This is analogous to the left hand side of \citep[eq. (7)]{Matveev/Trautman}); Matveev's second criterion, eq. (6), is the integrability condition for the scale connection. It does not play a role here.}  
Let us call $\tilde{T}$  the {\em generalized Matveev symbol} associated to a projective structure ${\mathfrak{p}}$  and a Weylian metric $[(g, \varphi)]$ subordinate to the conformal structure ${\mathfrak{c}}$. We thus get the result:
\begin{quote}
A projective structure ${\mathfrak{p}}$ and a conformal structure ${\mathfrak{c}} = [g]$ are  Weyl compatible, if and only if there is a real differential form $\varphi$ (not necessarily closed) such that  for some (and thus for any)  $\overline{\Gamma}\in \mathfrak{p}$ the generalized  Matveev  symbol (\ref{eq Matveev/Tratuman tensor}) vanishes,  $\widetilde{T}(g,\varphi,\overline{\Gamma}) =0$.
\end{quote}
 This argumentation follows step by step the one in \citet{Matveev/Trautman} although, as already said, this paper discusses the restricted case of Riemann compatibility and  $d\varphi=0$ only. It leads to a formal criterion for testing Weyl compatibility of a projective and a conformal structure. A   mathematical proof (or refutation) of the main result of EPS -- and with it of  the equivalence between Weyl compatibility and EPS-compatibility  -- remains a desideratum.   Trautman's remark in his commentary to the recent re-edition of the EPS paper has not lost its actuality:
 \begin{quote}
 I think it is clear that the subject requires and deserves further work. \citep[p. 1585]{Trautman:EPS}.
\end{quote}  


\subsection{How does the standard model relate to gravity?}
\label{subsec: SM and grav}
Two mutually related aspects of the standard model of elementary particle physics (SM) 
cry out for considering it from a Weyl geometric perspective if  one wants to know more about a possible link of the SM to gravity: the nearly scale covariance of the SM  Lagrangian (of weight -4) in a Minkowskian background with only the Higgs mass term breaking the scale symmetry, and the role of the Higgs scalar field $\Phi$ (weight $w(\Phi)= -1$) itself. 
Since the first decade of the SM this has motivated authors  to adapt the SM Lagrangian  to a Weyl geometric framework and to enhance it by a scale invariant gravity sector. In the first three decades that often happened without most authors knowing about similar attempts undertaken by other colleagues. Several of them brought forward idea that the (norm of the) initially still  hypothetical Higgs field might play the role of a gravitional scalar field which couples to the generalized Hilbert term of (\ref{L-grav general}) \citep{Smolin:1979,Cheng:1988,Drechsler:Higgs,Drechsler/Tann}.\footnote{Cf. \citep[sec. 11.5]{Scholz:2018Resurgence}.}

With the role of the Higgs field being established and elevated to the status of  an object of empirical research, all the more after the empirical confirmation of its physical reality in 2013, the attention has shifted to models with (at least) two scalar fields, a gravitational scalar field $\phi$ coupling to the Hilbert term and the Higgs field $\Phi$ as part of the matter sector, both scaling with the Weyl weight $-1$ respectively the mass/energy weight +1 used by the elementary particle physicists. This allows to consider models in which the Higgs field couples indirectly to the gravitational sector through  a common biquadratic potential term with the gravitational scalar field (\ref{L V4})  and/or directly (non-minimally)  by an additional term $\alpha_1' |\Phi|^2R$ added to the gravitational Lagrangian (\ref{L-grav general}). 
These  approaches are not mutually exclusive and may turn out crucial for establishing a link between the SM and gravity. 

Among the more recent studies some  explore the conceptual and formal framework for importing the SM into a Weyl geometric scalar tensor theory of gravity with  a modified Hilbert term in  the classical Lagrangian (without  quadratic gravitational terms and without considering quantization) \citep{Nishino/Rajpoot:2004,Nishino/Rajpoot:2007,Nishino/Rajpoot:2009,Scholz:2011Annalen,Scholz:Paving,Quiros:2013,Quiros:2014a,Cesare/Moffat_ea:2016}. Interesting features arise already on this level  if one assumes  a common biquadratic potential for the gravitational scalar field and the Higgs field:
\beqarr V(\Phi, \phi)  &=&\frac{\lambda_1}{4}\left(|\Phi|^2 - \frac{ \mu}{\lambda_1}\phi^2\right)^2+ \frac{\lambda}{4}\phi^4 \, , \qquad \lambda,\, \lambda_1 > 0 \, , \label{L V4}  \\
  &=&   \frac{\lambda_1}{4}|\Phi|^4-\frac{\mu}{2}|\Phi|^2\phi^2+ \frac{\lambda'}{4}\phi^4 \,   \qquad \lambda' = \lambda + \frac{\mu^2}{\lambda_1} \, ,  \nonumber \\
\mathcal{L}_{V} &=& -V(\Phi, \phi) \sqrt{|g|} \,  \nonumber
\eeqarr 
Such a potential is considered in \citep{Shaposhnikov/Zenh"ausern:2009,Shaposhnikov_ea:2009} for studying global scale symmetry in the SM on Minkowski space and has been imported to conformal theories of gravity in \citep{Bars/Steinhardt/Turok:2014}.\footnote{For the introduction to a Weyl geometric framework see \citep{Cesare/Moffat_ea:2016,Scholz:Paving}.} If the dynamical terms do not disturb the rest state of the potential minimum too strongly, there is a common minimum of the potential  with $(\partial_{|\Phi|}V,\partial_{\phi}V)=0$  also in the  locally scale symmetric framework. This ``binds'' the fields together in the sense of keeping them proportional, $\phi \sim |\Phi|$, whichever scale gauge is considered.  In particular the gauge in which the gravitational scalar field is constant, Einstein gauge, is the same in which the norm of the Higgs field, i.e. its expectation value, is scaled to a constant. I propose to call it  the {\em Higgs gauge}.

 Then {\em the Higgs  gauge and the Einstein gauge are identical} in the rest state of the scalar fields, 
  and the Lagrangian mass expressions  of the form $\mu\, |\Phi|^2$ acquire the  constant value $m^2 = \mu \,|\Phi_o|^2= \mu_e \mathit{v}^2$  as usual in the electro-weak theory ($ \mathit{v}\approx 246\, GeV$ the reference energy of the electroweak interaction). In the literature this is sometimes considered as a "generation" of mass (as though the mass terms did not exist already before) and the choice of Higgs gauge as a  ``breaking'' of the scale symmetry (as though the scale symmetry had to be given up in the light of the physically  preferred choice of the scale gauge). 

The latter perspective is unhappy because it obscures an  important point for rounding off the Weyl geometric framework. As the mass term of the electron $\mu_e |\Phi|^2$ scales with the norm of Higgs field $|\Phi|$ like the masses of all elementary  particles, it  becomes constant with the latter, say $m_e^2=\mu_e |\Phi_o|^2= \mu_e v^2$. In accordance with this the Rydberg constant $R_y$, important for the determination of atomic frequencies, scales (classically) with $m_e$, this allows to infer that time measurements which rely on atomic frequencies can be read off in the Higgs gauge = Einstein gauge.\footnote{$R_y= \frac{e^4 m_e}{2\hbar}=\frac{\alpha_f^2}{2}m_e c²$, where $e$ denotes the charge of the electron, $\alpha_f$ the fine structure constant. The energy levels $E_n$ in the Balmer series of the hydrogen atom are given by $E_n= R_y n^{-2}$;   the other atomic frequencies are influenced by the Rydberg constant similarly, although in a more complicated way.  
With field quantization $\alpha_f$ becomes dependent on the energy scale under consideration; the Higgs gauge as discussed here is therefore clearly a feature  of the classical theory emerging  from the quantum level at long ranges.  }
This leads to the observation: 

\begin{quote} \label{observation ast}
($\ast$) \quad If Weyl geometric gravity with a gravitational scalar field $\phi$ is linked to the SM via a biquadratic potential of $\phi$ and the Higgs field $\Phi$, we  expect that the Einstein gauge of the gravitational scalar field is identical to the Higgs gauge and corresponds to  measurement units according to  those of  the new SI conventions.\footnote{For the new SI conventions see footnote \ref{fn SI};  the connecting link to Weyl geometric gravity  is discusssed in  \citep[sec. 5]{Scholz:Paving}.}
 This can be interpreted as an adaptation of atomic clocks to the local field constellation of the pair $(\Phi,\phi)$, the Higgs field and the gravitational scalar field.
 \end{quote}
  
 From a general point of view this is close to Weyl's hypothesis of an adaptation of clocks to the local  constellation of the gravitational field. But because of the central role of the Higgs field in the adaptation to $(\Phi,\phi)$ the latter gains a  more solid physical foundation than  Weyl's rather  speculative  proposal relating to the scalar curvature of spacetime (sec.  \ref{Weyl clock adaptation}). Note also that  a biquadratic potential (\ref{L V4}) gives the Higgs field a distinguished relation among the matter fields to the gravitational sector. Th other way round, the gravitational scalar field $\phi$ enters the  so-called {\em Higgs portal} in a particularly simple form.

Present day physicists  want to understand the establishment of the Higgs gauge and/or Einstein gauge as  a  breaking of symmetry. This is what, e.g., the authors of \citep{Nishino/Rajpoot:2004} do. They consider the   scalar field $\phi$ acquiring a constant values as a physical process of symmetry breaking, related to a Goldston boson which gives mass to the Weyl field and to the Higgs field. This is assumed to happen at an extremely high energy level far beyond the electroweak scale. Other authors go a step further and discuss a possible  dynamical underpinning by a mechanism of a spontaneous breaking of scale symmetry in close analogy to the Coleman/Weinberg mechanism proposed in the  early 1970s for the breaking of the $U(1)$ symmetry in electroweak theory \citep{Ohanian:2016,Dengiz/Tekin:2011,Tanhayi_ea:2012,Ghilencea:2019JHEP}. 

These studies work with a quantization of at least  the scalar field and the Weyl field (the scale connection).  Dengiz/Tekin for example start from a  scale invariant Lagrangian including quadratic curvature terms in the gravitational sector. Applying  a perturbative quantization scheme for the gravitational scalar field they find a Coleman/Weinberg effect, after a ``tedious renormalization and regularization procedure'' at the two loop-level \citep[p. 4]{Dengiz/Tekin:2011}. Although there remains the caveat that at the minimum of the quantum corrected potential the perturbative calculation looses validity, they conclude that ``the Weyl symmetry of the classical Lagrangian will not survive quantization'' (ibid. p. 5). Ohanian and Ghilencea argue inversely. Ohanian assumes a scale invariant Lagrangian close to the Planck scale without the quadratic curvature terms, while Ghilencea uses a combined linear (modified Hilbert term) and quadratic gravitational Lagrangian with a prominent contribution of   Weyl's conformal curvature, which does not satisfy the Gauss-Bonnet constraint (\ref{Gauss-Bonnet constraint}).  Both authors find in their respective approaches  
 that a Coleman-Weinberg-like consideration for the scalar field leads to  breaking of scale symmetry for the low energy classical Lagrangian. This  reduces the geometry to a Riemannian framework and the gravity  to Einstein's theory.\footnote{\citep[p. 10f.]{Ohanian:2016}, \citep[p. 7]{Ghilencea:2019JHEP}}

 In a follow up paper to \citep{Dengiz/Tekin:2011} Tahanyi and the authors of the first paper come to a different conclusion. They undertake a more fundamental study of quantum properties of the Weyl geometrically  extended gravitational sector including quadratic expressions in the curvature and derive the full gravitational particle spectrum of the theory for an (Anti-) de-Sitter and a Minkowski background. They find the Gauss-Bonnet relation (\ref{Gauss-Bonnet constraint}) as a necessary constraint for unitarity of the quantized field theory with gravitational Lagrangian  (\ref{L-grav general}) 
 and arrive at the  conclusion:
\begin{quote}
\ldots , the only unitary theory, beyond three dimensions, among the Weyl-invariant quadratic theories is the Weyl-invariant Einstein-Gauss-Bonnet model [our Lagrangian (\ref{L-grav general}) constrained by (\ref{Gauss-Bonnet constraint}), ES] which propagates a massless spin-2 particle
as well as massive spin-1 and massless spin-0 particles \citep[p. 8]{Tanhayi_ea:2012} 
\end{quote}
This is a highly interesting result. 
The mentioned particles are, of course, the graviton, the boson of the Weyl field (scale connection), and the boson of the gravitational scalar field. In this model the Weyl boson is extremely heavy as usual, and {\em  the Weyl field has only  short range effects}; on larger scales it can be  represented classically by an integrable scale connection. {\em The scalar field}, on the other hand (and different from Ohanian's model), {\em  can support long range gravitational interactions} in addition to the  ones mediated by the graviton. 
For the classical theory this boils down to a Lagrangian  constraint
\beq L_f= \lambda_{\mu \nu} f^{\mu\nu} \, , \label{Lagrange constraint f=0}
\eeq 
with Lagrangian multiplier functions $\lambda_{\mu \nu}$  enforcing $\alpha_2=0$ in (\ref{L-grav general}). 
 In the light of Tanhanyi et al.'s derivation the classical Einstein gravity limit of Ohanian and Ghilencea seems to  result from the special choices of their gravitational Lagrangians, both violating the Gauss-Bonnet constraint.

The question still remains, whether for a Weyl symmetric bare  Lagrangian -- a kind of classical template for a quantum field theory under construction -- the local scale symmetry is broken under quantization. Often the so-called {\em trace anomaly} is considered as an indicator that this is in fact the case. This means that the energy-momentum tensor of (pseudo-) classical quantum matter fields, represented by  scalar fields or Dirac spinors, has a vanishing trace, whereas  the expectation value of the quantized  trace no longer vanishes. The trace anomaly  has  puzzled theoretical physicists for a long time, and is often interpreted as implying the breaking of  scale invariance at the quantum level. 

The authors of \citep{Codello_ea:2013} come to a different conclusion. The trace anomaly is, of course,  present also in their approach, but  it {\em no longer signifies  breaking of the local scale invariance}. The reason  lies in a  cancellation of the trace  terms of the quantized fields by corresponding counter-terms that arise from a {\em complex} gravitational  scalar field which they call a ``dilaton''. In order to achieve this the authors use a coherent scale-covariantization of all terms by substituting dimensional constants with the appropriate power of the scalar field and a dimensionless coefficient. In a series of examples they show that ``with a suitable quantization procedure, the equivalence between conformal frames can also be maintained in the quantum theory'' \citep[p. 21]{Codello_ea:2013}. They even can show that the renormalization flow preserves the Weyl symmetry  in their case studies.  This stands in   contrast to Dengiz/Tekin's observation mentioned above.  It would be interesting to see,  whether a coherent scale covariantization as used by Codello/Percacci et al. can be implemented in their scheme too, leading to a maintenance of the scale symmetry also in the context of a perturbative quantization of gravity.

\subsection{Open questions in cosmology and dark matter}
\label{subsec: cosmology and dm}
 Weyl geometric gravity can also be used to explore  the open problems in cosmology and dark matter phenomena from its sight. 
Already  Weyl started to consider  cosmological questions in his extended  geometrical framework. Since the retake of it by Dirac and the Japanese authors the conjecture  that the Weyl field, or now also the gravitational scalar field, might help understanding  dark matter effects has been expressed  by different authors time and again. This is so still today. There  even is an  author of the 1970/80s, A. Maeder, who  has taken up  the line of investigation following Dirac and Canuto  after an interruption of several decades, with the goal of establishing links  to the present research in cosmology and astrophysics  \citep{Maeder:2017LCDM,Maeder:2017dm,Maeder:2017CMB}  (cf.  sec. \ref{subsec: WST}). This is an interesting attempt, but as I can see  the approach of these papers suffers from an incoherent usage of scale transformation conventions and an unclear dynamical basis. It needs more time and astronomical expertise to judge whether the striking results claimed by the author with regard to accounting for recent empirical data (CMB, galactic rotation, cluster dynamics) are crucially affected by the theoretical deficiencies, or whether they can be upheld  in an improved framework. If so, Maeder's approach would indicate a blueprint for a  unifying account of cosmology and dark matter phenomena,  which is  rather, perhaps even fundamentally, different from the present mainstream picture of cosmology and dark matter.

In the sequel I discuss  two less incisive examples of recent research approaches to the field from the point of view of Weyl geometry. A review of the cosmological investigations of the Brazilian group initiated by M. Novello and its external Greek contributor J. Miritzis (see sec. \ref{subsec: WST})  has been given at another place.\footnote{\citep[sec. 11.6.4]{Scholz:2018Resurgence}} 
 Here I add a short discussion of  studies by a group of authors around J. Jim\'enez and T. Koivisto (later also including L. Heisenberg) \citep{Jimenez/Koivisto:2014,Jimenez/Koivisto:2016,Jimenez/Heisenberg/Koivisto:2016} and  supplement it by a  remark on a  recent proposal for modelling galactic and cluster dynamics in a Weyl geometric scalar tensor theory with an unconventional kinetic term \citep{Scholz:2019WST}. A comparison of the Weyl geometric approach(es) with  conformal cosmological models  of a more general type, in particular C. Wetterich's   ``universe without expansion'' \citep{Wetterich:1988,Wetterich:2013}, would be an interesting task, but cannot be carried  out here.

Let us start the report on  the  first example with an interesting remark made by the two  authors of the first paper about the reason for  studying a Weyl geometric framework:
\begin{quote}
\ldots  (T)he nearly scale invariant spectrum of cosmological perturbations
and the fact that the Higgs mass is the only term breaking scale invariance in the Standard Model of elementary particles are very suggestive indications that this could be an actual symmetry of a more fundamental theory that has
been spontaneously broken \citep[p. 2]{Jimenez/Koivisto:2014}.
\end{quote}
The  paper is intended to lay  the ground for further investigations of gravity and cosmology in the framework of Weyl geometry.

In the second paper,  \citep{Jimenez/Koivisto:2016},   the framework is extended to what the authors call a {\em generalized Weyl geometry}. Here a   linear connection $\Gamma$  with non-vanishing torsion is accepted, accompanied by  a concomitant generalization of  the Weyl geometric compatibility condition (\ref{compatibility metric}) between the Riemannian component of the Weylian metric and the covariant derivative $\nabla(\Gamma)$:
\beq \nabla \, g = - 2 b\, \varphi \otimes g \, , \label{generalized metric compatibility}
\eeq  
where  $b$ is any real valued parameter. 
The permissible  linear connections in any scale gauge $(g,\varphi)$ are then given by the  2-parameter family:
\beq \Gamma^{\mu}_{\nu \lambda} = \Gamma(g)^{\mu}_{\nu \lambda} + b' \delta^{\mu}_{\nu}\varphi_{\lambda} +  b \,\delta^{\mu}_{\lambda}\varphi_{\mu} - b' \varphi^{\mu}g_{\mu \nu}
\eeq 
For $b\neq b'$ the connection has torsion, with $\Gamma^{ \lambda}_{\mu\nu}-\Gamma^{\lambda}_{\nu\mu} $ (the {\em distorsion}) being a linear expression in the Weyl field $\varphi$.
For both parameters equal 1  the usual (torsion free) Weyl geometric structure is included in the generalized framework.\footnote{The authors start from a slightly more general linear connection with three parameters $b_1, b_2, b_3$  but soon specialize to the generalized metric compatibility of the general form (\ref{generalized metric compatibility}), which implies  $b_3=2b_1-b_2$. The corresponding  parameters used here are then given by $b=(b_1-b_2), \, b' =\frac{1}{2}(b_2+b_3)  $.}

 The authors study quadratic gravity theories in this generalized framework.   With an interesting argument they constrain the quadratic terms by the (external) condition that their Riemannian component reduces to the Chern-Gauss-Bonnet form,  (\ref{Gauss-Bonnet constraint}) for case $n=4$:
\begin{quote}
The reason for this restriction is the well-known fact that the Gauss-Bonnet term (\ldots) leads to second order equations of motion for the metric tensor and, thus, avoids the potential presence of Ostrogradski's instabilities \citep[sec. III]{Jimenez/Koivisto:2014}.
\end{quote}
This adds a classical field theoretic argument to the quantum field consideration reproduced in sec. \ref{subsec: SM and grav}  for constraining  quadratic gravity theories by the Gauss-Bonnet relation.  

Like the Brazilian group Jim\'enez/Koivisto 
  don't  insist on a  scale invariance of their complete Lagrange density. Vaguely appealing to some ``spontaneously''  breaking of the scale symmetry at high energies, they  add a  (Weyl geometric) Einstein term which does the job, or at least documents a  broken scale symmetry: 
\[\mathfrak{L}_E= -\frac{1}{2} M_p^{n-2}R \sqrt{|\mathit{det}\,g|} \qquad  (n= \mathit{dim}\, M)
\]

 In later papers they use more general considerations also violating the scale invariance of the total Lagrangian. This seems to be a precondition for  constructing cosmological models which allow to establish links with the present mainstream discourse. \citep{Jimenez/Koivisto:2016}   starts with  a general functional term of the form $f(R)$ (with Weyl geometric scalar curvature $R$), which is transformed via a Lagrangian constraint trick into the form of a scalar tensor theory of  Brans-Dicke type, with the Lagrangian constraint function playing  the role of the scalar field \citep[eq. (21)]{Jimenez/Koivisto:2016}. In very simple cases (``taking into account only the leading order quadratic correction to the Einstein-Hilbert  action'',  $L= R+ \frac{R^2}{6M^2}$) the authors arrive at a model which in the Riemannian case implies a special type of inflationary scenario (``Starobinsky inflation''). In their Weyl geometric generalization they find a 1-parameter generalization of this model due to the Weylian scale connection, similar to a vector field introduced also by other approaches in the literature. 
 
 This is only the starting point for more extensive studies of cosmological models in  \citep{Jimenez/Heisenberg/Koivisto:2016} undertaken with extended (wo-)manpower by including L. Heisenberg into the team. The three researchers study a class of quadratic gravity theories  in generalized Weyl geometry satisfying the  Gauss-Bonnet relation and with  a scale symmetry breaking Einstein-Hilbert term.  By assuming an external constraint  of a  purely temporal and homogeneous scale connection they reduce  their geometrical ``vector tensor'' theory to a scalar tensor theory with a scalar field (the temporal component of the scale connection) satisfying a non-dynamical algebraic constraint (ibid. p. 8). They study perturbations around the de Sitter universe and look for bouncing solutions, which arise for non-realistic values of the paramaters only. Finally properties arising from adding a  matter component to   the Lagrangian are studied.
 
  Resuming their work the authors find that
 \begin{quote}
\ldots  the class of vector-tensor theories 
studied in this work and which naturally arise in the framework of geometries with a linear vector distortion can give rise to a rich an[d] interesting phenomenology for cosmology. \citep[p. 23]{Jimenez/Heisenberg/Koivisto:2016}
 \end{quote}
 This adds rich material to the stock of models already studied without generalizing the Weyl geometric structure by the Brazilian group and Miritzis; but it keeps still far away from the quest for an improved, or even alternative, understanding of empirical  data.
  From a cultural point of view one might get the impression that now also part of the work in Weyl geometric gravity has joined the  ``postmodern'' wave of studying the widest possible range of theoretical models with only  vague allusions to the empirical data collected in astronomy and observational cosmology.
  
The approach in \citep{Scholz:2019WST}  has a different motivation. It deals with  the question whether  a Weyl geometric linear  gravity theory ($ \beta_1 = \beta_2 =\beta_3 =0$ in (\ref{L-grav general})) with scalar field can reproduce the deviation of galactic dynamics from the Newtonian expectation, i.e. imply MOND-like dynamics for low velocity trajectories, {\em and lead to a  gravitational light deflection} which  one would expect for a dark matter explanation of the anomalous galactic dynamics.\footnote{MOND stands for the {\em modified Newtonian dynamics} devised by M. Milgrom \citep{Milgrom:1983}. It is characterized by a new constant $a_0$ with the dimension of an acceleration. For a recent survey see \citep{Famaey/McGaugh:MOND}. } 
According to (\ref{Lagrange constraint f=0}) the author assumes a framework of  integrable Weyl geometry (external constraint $f_{\mu \nu} =0$ in (\ref{L-grav general})) and studies the consequences of an  unconventional kinetic term $L_{\partial \phi}$ of the scalar field. So far this approach can be understood as a generalized BD-type scalar field theory in a Weyl geometric setting;  it furthermore has to assumes  a screening of the scalar field which restricts its sphere of action to regimes of extremely weak gravity like in the superfluid approach to anomalous galactic  dynamics \citep{Berezhiani/Khoury:2015,Berezhiani/Khoury:2016,Berezhiani/Khoury:2019}. The kinetic term of the scalar field combines  a cubic expression $L_{\partial \phi3}$  in the partial derivatives of the scalar field  $|\partial \phi |^3$  and a  second order derivative (in $\partial^2\phi$) term $L_{\mathit{\partial 2\phi}}$, both of weight $-4$  and a scale invariant  Lagrange density,
\beq  L_{\partial \phi} = L_{\partial \phi3} + L_{\mathit{\partial 2\phi}} \, .
\eeq 
  $L_{\partial \phi3}$ is similar to the kinetic term in the first relativistic generalization of the MOND theory introduced by J. Bekenstein and M Milgrom, called RAQUAL (``relativistic aquadratic Lagrangian'') \citep{Bekenstein/Milgrom:1984}. $L_{\mathit{\partial 2\phi}} $ is adapted from the cosmological studies of Novello and the Brazilian group  \citep{Novello/Oliveira_ea:1993}, but has been modified such that  $w(L_{\mathit{\partial 2\phi}})=-4$. Due to its special form it keeps the dynamical equation for $\phi$  of second order.\footnote{For the   variation with regard to $\phi$ the second order terms of  $L_{\mathit{\partial 2\phi}}$  form a divergence expression; not so, however, while  varying  $g$.  Thus the dynamical equation of $\phi$ remains of second order, while  the $L_{\partial 2\phi}$  term contributes essentially to the source term of the Einstein equation (the Hilbert energy-momentum tensor).}
  
The Lagrangian density of baryonic matter is assumed to be brought into a scale invariant form such that its (twice covariant) Hilbert energy-momentum tensor scales with weight $-2$. Stretching the result of the Geroch-Jang theorem slightly,  from test bodies to extended bulk of matter, the free fall trajectories of matter can reasonably be assumed to follow Weyl geometric geodesics  (see end of sec. \ref{subsec: Weyl geometric gravity}). 
The Einstein gauge is considered as distinguished for indicating measured values of observable quantities (cf.  ($\ast$) in section  \ref{subsec: cosmology and dm}).  The Levi-Civita part $\Gamma(g)$ of the affine connection in Einstein gauge can be computed from the Riemannian component of the Weyl geometric Einstein equation, and the scale connection contribution $\Gamma(\varphi)$  from the scalar field equation. 
 Step by step   the following properties of the model are established  in \citep{Scholz:2019WST}:
    \begin{itemize}
    \item[1] With the scale connection in Einstein gauge written in the form  $\varphi \underset{Eg}\doteq  d \sigma$ like in (\ref{eq Einstein gauge}), the  weak field approximation of the scalar field equation leads to a non-linear Poisson equation for $\sigma$, known from Milgrom's theory for the deep MOND regime,
\[ \nabla \cdot (|\nabla \sigma|\nabla \sigma)= 4\pi G\, a_o\, \rho_{\mathit{bar}} \, 
\] 
 with the usual MOND constant $a_0$;  Scholz calls it the {\em Milgrom equation}.  It is  sourced by  baryonic matter only.\footnote{In complete generality, the scalar field equation in Einstein gauge  is a scale covariant generalization of the Milgrom equation.} 
    \item[2] $\sigma$ induces  an acceleration   for free fall trajectories, due to the scale connection,  $ a_{\varphi} = -\nabla \sigma$. It is  of deep MOND type. 
    \item[3] The weak field approximation for determining the Riemannian component $g$ of the Weylian metric in the Einstein gauge  boils down to a  Newton approximation exactly like in  Einstein gravity. It is sourced by the baryonic matter only, not by the scale field (see 4), and induces an acceleration $a_N$ like in Newton gravity with potential $\Phi_N$.
 
  \item[4] The scalar field has an  energy momentum tensor with non-negligible energy density   $\rho_{\phi}= (4\pi G)^{-1}\,\nabla^2\sigma$. This looks like dark matter,  but is has  negative pressures  $p_1, p_2, p_3$ with peculiar properties: \\
 (i)  The contribution of the scalar field  to the Newton-Poisson source (of the Einstein weak field equation) vanishes, $\rho_{\phi}+\sum p_j=0$. \\
 (ii) On the other hand it contributes   to the gravitational light deflection. In the central symmetric case it does so  by  an  additive contribution of  $2 \sigma$ to the deflection potential. For  baryonic matter in Einstein gravity the deflection potential is $2\Phi_N$.\footnote{The deflection angle  is computed from the first two components of the Ricci tensor like in 
  \citep[p. 288f.]{Carroll:Spacetime}. It is the gradient of $\frac{1}{2}(-h_{00}+h_{jj})$ (for any $1\leq j\leq 3$), where $h$ is the deviation of a weak field metric from the Minkowski metric,  $g=\eta + h$. $\frac{1}{2}(-h_{00}+h_{jj})$ thus plays the role of a {\em deflection potential}.}
  \item[5] (i)  2 and 3 together imply a MOND-like phenomenology for test particles, with  a deep MOND acceleration  added to the Newtonian one, both with the source $\rho_{bar}$, in the regime of activity of the scalar field (screened for larger accelerations).\\
   (ii) The light deflection of the model is  identical to the one expected  for a source given by $\rho_{\mathit{bar}}+\rho_{\phi}$  in Einstein  gravity.\\  
  (iii) The ``dynamical phantom mass'' $\rho_{\mathit{phant}}$, i.e. the mass density expected in Newton gravity for generating  the  additional acceleration $a_{\varphi}$ predicted by  the model, 
  coincides with the ``optical phantom mass'' $\rho'_{\mathit{phant}}$ which one would expect in a dark matter approach for the same light deflection as  in the model.  Both agree with the    
   mass/energy density of the scalar field,   $\rho_{\phi}=\rho_{\mathit{phant}}=\rho'_{\mathit{phant}}$. 
  \end{itemize}
Resum\'ee: The {\em gravitational effects} for test bodies and for light rays of the model's scalar field $\phi$, observed  in the Einstein gauge, {\em coincide with those of a halo of (pressure-less) dark matter} generating a  Newtonian potential $\sigma$, although the energy-momentum of the scalar field  carries    negative pressures and is, in this sense, similar to ``dark energy''. The potential  $\sigma$ of the additional acceleration due to the scale connection is given by a  Milgrom equation (with  the baryonic matter  as its source). 

This is an exceptional model among the many models which  propose relativistic generalizations of Milgrom's modified Newtonian dynamics. Most of them work with several scalar fields, often also with an additional dynamical vector field, and/or strange (non-conformal) deformations of the metric. The original RAQUAL theory, with only one additional scalar field, is inconsistent with the observed amount of  gravitational light deflection in galaxies and clusters. To my knowledge the present Weyl geometric scalar tensor model is the only one  with a {\em  single  scalar field} (at large scales, thus not counting the Higgs field here) leading to {\em MOND-like dynamics and an according enhancement of the gravitational light deflection}.\footnote{Of course that has to be checked by astronomers. Heuristic considerations indicate that the model may even ``predict'' the dynamics of galaxy clusters without assuming  additional dark matter  \citep[sec. 3.2]{Scholz:2019WST}.} 
But alas, its kinetic term looks so strange that it may remain an exercise in theory construction only. In any case, it deserves to be mentioned here as an outcome of  recent and even  present  investigations in Weyl geometric gravity.

\section{An extremely short glance  back and forth}
\label{sec: Looking back and forth}
We have seen an eventful trajectory of Weyl's proposal for generalizing Riemannian geometry during the last century: from a promising first attempt (1918 to the middle of the 1920s) at formulating a type of geometry which was intended to bring the framework of gravity closer to the principles of other field theories, at that time essentially electromagnetism, through a phase of disillusion and non-observance (1930 -- 1970), to a new rise in the 1970s. In the last third of the 20th century dispersed contributions to various fields of physics  brought Weyl geometry back as a geometrical  framework for actual research, partially fed by the introduction of a hypothetical scalar field which had at least one foot in the gravitational sector. But these new researches were far from influential on the main lines of development in theoretical physics.

 Although first hypothetical links between the new attempts at Weyl geometric generalizations of gravity theory and the rising standard model of elementary particle physics were soon formulated, they remained rather marginal. Also the vistas opened up by Weyl geometric scalar tensor theories for astronomy at large scales and cosmology remained without a visible impact on the respective communities.  One reason for the overall reservation may have been the fact  that, at that time, scalar fields played only a hypothetical, perhaps even only  formal role within physical theories.  
 This changed, of course, with the empirical confirmation  of the Higgs boson.
 
  It may be worthwhile to investigate whether an additional scalar field with links to the Higgs portal and gravity can help to answer   some of the open questions of the present foundational research in physics. Some of the  recent work surveyed in this paper make already some steps in this direction.  As it is always difficult to discern reliable patterns of actual developments,  I leave it as a question for the future to judge whether this is a fruitful enterprise, or  just  another contribution to  screening possible theory alternatives without clear evidence of empirical support.
  
   To end with a personal remark as a historian of mathematics who has spent some time on studying the development of  this  subfield of physical geometry: it is pleasant to see that, contrary to Weyl's disillusionment in this regard after 1920, his beautiful and conceptually convincing idea for generalizing Riemannian geometry is still alive and helps triggering new lines of research also a century after it was first devised.

\newpage
\small
 \bibliographystyle{apsr}
 \bibliography{%
/home/erhard/Dropbox/Datenbanken/BibTex_files/lit_hist,%
/home/erhard/Dropbox/Datenbanken/BibTex_files/lit_mathsci,%
/home/erhard/Dropbox/Datenbanken/BibTex_files/lit_scholz,%
/home/erhard/Dropbox/Datenbanken/BibTex_files/lit_Weyl,%
/home/erhard/Dropbox/Datenbanken/BibTex_files/lit_Einstein}
\end{document}